\documentclass[aps,pra,twocolumn,amsmath,amssymb,amsfonts,superscriptaddress,floatfix,nofootinbib]{revtex4-1}
\usepackage{epsfig} \usepackage{graphicx}
\usepackage[pdftex,bookmarks]{hyperref}
\usepackage[pdftex]{color}
\usepackage[table]{xcolor}  

\newcommand{\bgar}{\begin{eqnarray}}
\newcommand{\enar}{\end{eqnarray}} 
 
 \newcommand{\be}{\begin{equation}}
\newcommand{\ee}{\end{equation}}  
 \def\mincirc{\lower
  3pt\hbox{$\buildrel<\over{\hbox{$\mathchar"218$}}$}}
\newcommand{\eotvos}{E$\ddot{\rm o}$tv$\ddot{\rm o}$s}
\newcommand{\eotwash}{E$\ddot{\rm o}$t-Wash}

\begin{document}

\title{\bf Relevance of the weak equivalence principle and experiments to test it: \\ lessons from the past and improvements expected in space\\}

 \date{\today}

\author{Anna M. Nobili}
\affiliation{
Dept. of Physics ``E. Fermi'', University of Pisa, Largo B. Pontecorvo 3,
56127 Pisa, Italy}
\affiliation{INFN-Istituto Nazionale di Fisica Nucleare, Sezione di Pisa, Largo B. Pontecorvo 3,  56127 Pisa, Italy}

\author{Alberto  Anselmi}
\affiliation{Thales Alenia Space Italia, Strada Antica di Collegno 253, 10146 Torino, Italy}

\begin{abstract}
Tests of the Weak Equivalence Principle (WEP) probe the foundations of physics. Ever since Galileo in the early 1600s, WEP tests have attracted some of the best experimentalists of any time. Progress has come in bursts, each stimulated by the introduction of a new technique: the torsion balance, signal modulation by Earth rotation, the rotating torsion balance. Tests for various materials in the field of the Earth and the Sun have found no violation to the level of about 1 part in $10^{13}$. A different technique, Lunar Laser Ranging (LLR), has reached comparable precision. Today, both laboratory tests and LLR have reached a point when improving by a factor of 10 is extremely hard. The promise of another quantum leap in precision rests on experiments performed in low Earth orbit. The Microscope satellite, launched in April 2016 and currently taking data, aims to test WEP in the field of Earth to $10^{-15}$, a 100-fold improvement  possible thanks to a  driving signal in orbit almost $500$ times stronger  than  for torsion balances on ground.  The `Galileo Galilei' (GG) experiment, by combining the advantages of space with those of the rotating torsion balance, aims at a WEP test 100 times more precise than Microscope, to $10^{-17}$. A quantitative comparison of the key issues in the two experiments is presented, along with  recent experimental measurements relevant for GG. Early results from Microscope, reported at a conference in March 2017, show measurement performance close to the expectations and confirm the key role of rotation with the  advantage (unique to space) of rotating the whole spacecraft.   Any non-null result from Microscope would be a major discovery and call for urgent confirmation; with 100 times better precision GG could settle the matter and provide a deeper probe of the foundations of physics.
\end{abstract}

\maketitle

\section{Introduction}
\label{Sec:Introduction}

The General theory of Relativity (GR)~\cite{Einstein1916GTR} stands on the fundamental assumption that in a gravitational field all bodies fall with the same acceleration regardless of their mass and composition, a `fact of nature' known as the Universality of Free Fall (UFF) or the Weak Equivalence Principle (WEP). 
The WEP has been tested for various materials in the field of the Earth and the Sun, and no violation has been found to the level of about  $\Delta a/a\simeq10^{-13}$~\cite{TBfocusIssue2012}   ($\Delta a$ is the difference in acceleration between two test bodies falling in a gravitational field with mean acceleration $a$, referred to as the `driving signal').

A WEP experiment is both a test of the foundation stone of GR and a search for a new long range field coupling to matter in a way that depends on composition. A confirmed violation would have the same significance as the discovery of a new force of nature. There is no firm prediction as to the level at which the violation should occur. However, the WEP is so fundamental a postulate that any experiment that can push limits by many orders of magnitude is highly significant, whether it finds an effect or not. 

Substantial progress in WEP test precision has always depended on the introduction of a new technique:
 the torsion balance at the turn of the 20th century (\eotvos), the Sun providing a daily  modulated signal source  (Dicke, Braginsky in the 1960s-70s), the rotating torsion balance (Adelberger and collaborators, from the early 1990s to this date). Nowadays, laboratory experiments have run their gamut and any further progress is small and comes at slow pace. 
A completely different technique, Lunar Laser Ranging (LLR), tests the WEP for the Earth and Moon as bodies of different composition falling in the field of the Sun. Such tests have reached a precision similar to the torsion balance~\cite{LLRWilliams2012,LLRMueller2012} and a $10$-fold improvement requires not only mm-level laser ranging, but also a matching improvement of the physical model  which describes the Earth-Moon system~\cite{APOLLOfocusIssue2012}.

Today, only space experiments seem capable of a significant step forward. Just because in orbit the driving signal from Earth is stronger by a factor of almost 500 than it is for the torsion balance on ground, a carefully designed orbiting experiment can target a precision improved by a similar  factor. 

The Microscope mission, launched in April 2016 into a low altitude, sun-synchronous orbit, aims to test WEP in the field of  Earth to $10^{-15}$~\cite{MicroscopeFocusIssue2012}, a 100-fold improvement over the rotating torsion balances which can be achieved with a lower sensitivity to differential accelerations  thanks to the stronger driving signal in orbit.


The guiding principle of the `Galileo Galilei' (GG) small satellite mission~\cite{GGfocusIssue2012} is to fully exploit the advantages of space as well as those of the rotating torsion balance,  so as to design a balance  optimized for testing the WEP at zero-$g$.  GG aims to reach $10^{-17}$:  a four order of magnitude improvement over the best ground experiments and a 100-fold improvement over Microscope. For GG to achieve its target, it must be about 20 times more sensitive to differential accelerations than rotating torsion balances, which is possible by exploiting weightless conditions inside  an isolated  co-rotating laboratory (the spacecraft)  passively stabilized by  rapid  $1\,\rm Hz$ rotation around the symmetry axis.

Today, experimental evidence pointing the way out of the current physics impasse is hard to find, and even a hint of an effect from Microscope would cause excitement and call for confirmation by new measurements with increased precision. GG could provide validation at the level of $1\%$ and settle the matter. Recently, GG was proposed as a candidate in the European Space Agency's M5 competition for a new medium-sized science mission, and is awaiting further inquiry, having passed the first round of selection.

The paper is organized as follows. 

Sec.\,\ref{Sec:WEPrelevance}     presents the WEP  as an `experimentum crucis' of modern physics. 
Sec.\,\ref{Sec:GalileoToEotvos} presents the principles of the torsion balance experiment introduced by \eotvos\ for testing the equivalence between inertial and gravitational mass,  a decisive progress over previous experiments with pendulums by Galileo, Newton, Bessel and others.  
Sect.\,\ref{Sec:EotvosToEotwash} elaborates on rotation as the other key element for increased precision. 
Sec.\,\ref{Sec:StateOfArt} summarizes the state of the art of  current WEP experiments and their limitations. 
Sec.\,\ref{Sec:GroundToSpace} makes the case for a space  experiment in low Earth orbit as the way out of such limitations. 
Sec.\,\ref{Sec:Microscope} discusses the key features of the Microscope space experiment aiming at $10^{-15}$. 
Sec.\,\ref{Sec:GG} shows how, on a similar orbit as Microscope and without requesting cryogenic temperatures, a different experiment design  allows GG to aim at a  100 times better precision, to $10^{-17}$.  Recent experimental results relevant to the GG mission are also reported, along with positive news from the Microscope orbiting experiment which corroborate the choice of exploiting  rotation in space. 
Sec.\,\ref{Sec:Conclusions} draws the conclusions.


\section{An `experimentum crucis' of modern  physics}
\label{Sec:WEPrelevance}


The UFF was established experimentally by Galileo at the turn of the 17th century using two pendulums of different composition (see~\cite{NobiliAJP2013} for a general discussion on the universality of free fall and the equivalence principle). In 1687, in the opening paragraph of the `Principia', Newton stated the equivalence of inertial and gravitational mass and then went on to derive the equations of motion showing that all masses fall with the same acceleration under the gravitational attraction of the Earth. If inertial and gravitational mass are equivalent, UFF holds: this was the `equivalence principle' until the early 20th century. 

In 1907, Einstein made the crucial leap from Newton's principle (now referred to as the weak equivalence principle, WEP), to the strong equivalence principle, SEP (also referred to as the Einstein Equivalence Principle, EEP). In the words of Robert Dicke~\cite{DickeLesHouchesLectures}:  {\it ``The strong equivalence principle might be defined as the assumption that in a freely falling, non-rotating, laboratory the local laws of physics take on some standard form, including a standard numerical content, independent of the position of the laboratory in space and time. It is  of course implicit in this statement that the effects of gradients in the gravitational field strength are negligibly small, i.e. tidal interaction effects are negligible. \ldots this interpretation of the equivalence principle, plus the assumption of general covariance is most of what is needed to generate Einstein's general relativity. ''} 
Should experiments invalidate UFF (and the WEP), they would invalidate the SEP as well.

 As experimental evidence for UFF, Einstein took the results of of \eotvos\ (\cite{Einstein1916GTR}, p.\,773), who had achieved an impressive 1000-fold improvement over previous experiments by suspending test masses of different composition on a torsion balance rather than individual pendulums.

The Standard Model of particle physics and the General theory of Relativity, taken together, form our current view of the physical world. While the former governs the physics of the microcosm, the latter governs physics at the macroscopic level. Gravity couples in the same way to all forms of mass-energy, in all bodies, regardless of composition. Such universal coupling makes gravity different from all known forces of nature described by the Standard Model, and is at the heart of the fact that the two theories have so far resisted all attempts at reconciliation into a single unified picture of the physical world. This is the crossroad physics faces at the present time, which is of vital interest not only to theorists, especially given that the nature of about $95\%$ of the matter-energy in the Universe --the so called dark matter and dark energy-- is presently unknown.

An experiment capable of testing UFF to extremely high precision can potentially break this deadlock. The situation is reminiscent of that at the end of the 19th century, when Michelson and Morley tested by very precise light interferometry the propagation of the newly discovered electromagnetic waves through the hypothetical ether~\cite{MM1887}. Their null experimental result showed beyond question that although its existence was generally assumed, there was in fact no ether; which led to the special theory of relativity. While Michelson and Morley knew which precision their interferometer had to achieve in order to detect the relative velocity between the Earth and the ether, we do not know which precision a test of UFF-WEP should reach to detect a violation, if any. Nonetheless, the issue is so important and the potential reward so huge that many prominent experimental physicists have spent long years in such tests, renewing the effort whenever the possibility for an improvement has arisen. 

A WEP experiment is both a test of the foundation stone of GR and a search for a new long-range field coupling to matter in a way that depends on composition (phenomenologically, on powers of the atomic number $Z$ and nucleon number $A$). The mass-energy content ($A/Z$ ratio; electromagnetic effects in the proton mass, in the neutron mass and in the binding energy of the nucleus; etc.) varies greatly in different atoms and the validity of WEP at very high precision --implying that all forms of mass-energy fall with the same acceleration-- is a very strong constraint for all physical theories to comply with~\cite{DickeLesHouchesLectures}. In the years leading to GR, Einstein realized that the theory could stand or fall depending on the results of  a single experiment, and even went so far as proposing one himself, calling it a `simple experiment which would have the significance of an experimentum crucis'~\cite{IllyEinstein}.

Gravitational self-energy, neutrinos and photons, matter and antimatter, in the purely geometrical treatment  of General Relativity, all obey the WEP. Conversely, a confirmed violation could provide the so far missing clue to a more comprehensive physical theory. The higher the precision of the test, the higher the chances to find new physics. 

WEP tests are null experiments, by their nature among the most precise types of experiments in physics. They are conceptually simple, can rely on well proven techniques of experimental physics and do not require large apparata or resources, which makes it easier to detect and  control systematic errors. Their very high probing power has already been demonstrated in the lab to $10^{-13}$ (Sec.\,\ref{Sec:StateOfArt}); however a quantum leap in precision can only be achieved by performing the experiment in space  (Sec.\,\ref{Sec:GroundToSpace}).  

%

 \section{From Galileo to E\"OtV\"OS:  replacing pendulums with the   torsion  balance}
\label{Sec:GalileoToEotvos}

Galileo tested the UFF using  masses of different composition suspended from wires of the same length and checking how long the two pendulums  would keep on step with each other~\cite{TorreSTEPsymposium}. From Galileo's own description (see~\cite{GalileoDiscorsi}, pp.\,128-129) and based on a modern analysis of his experiments~\cite{FuligniIafollaGalileo1993}, it is apparent that he achieved a test  of the universality of free fall to about $10^{-3}$.  Newton reports his own pendulum experiments in the `Principia' and concludes that inertial and gravitational mass are equivalent  with a similar precision. Pendulum tests were later improved to reach  a few $10^{-5}$.
\begin{figure}[h!]
\begin{center}
\includegraphics[width=0.50\textwidth]{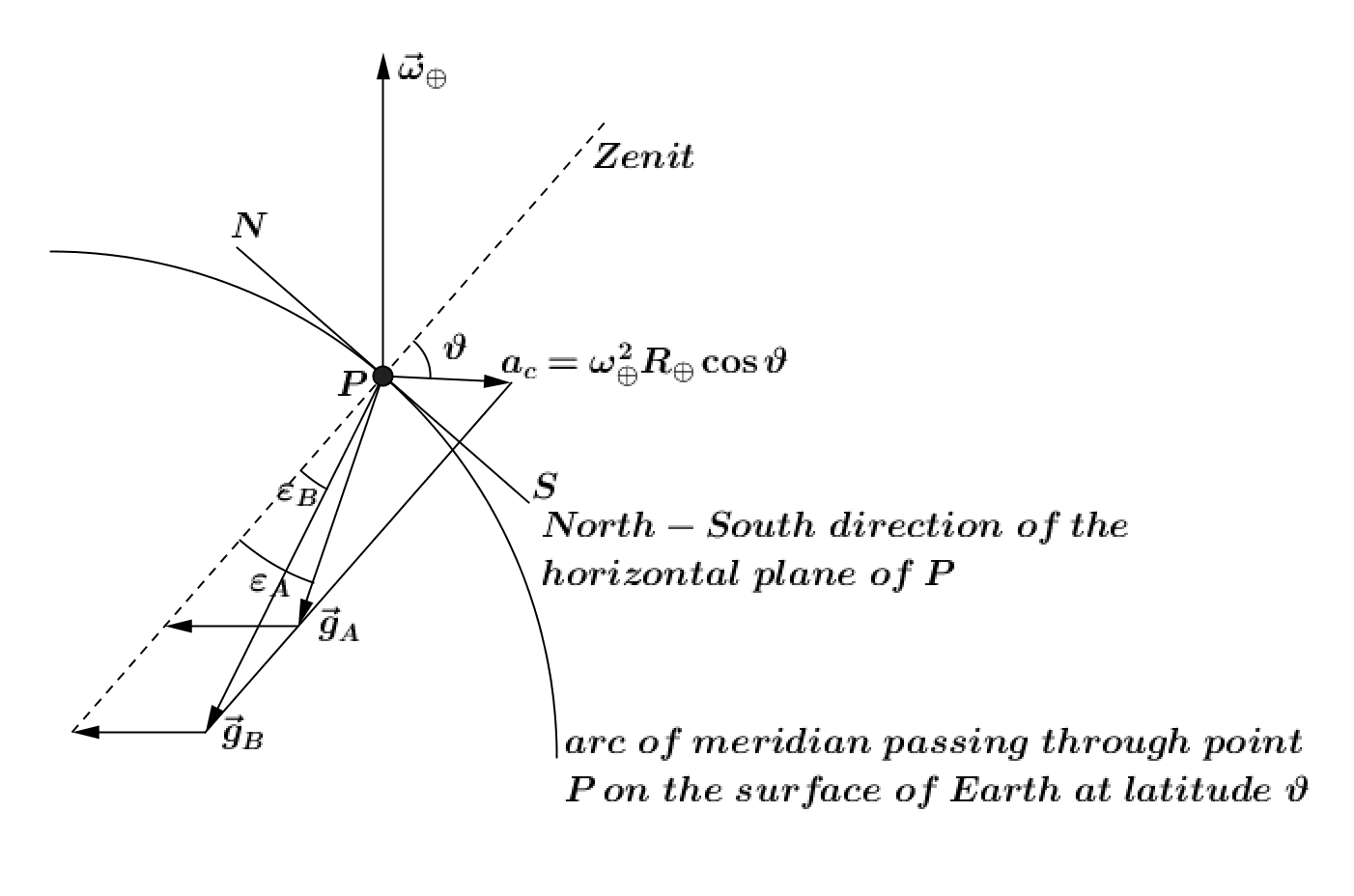} 
\caption{Two plumb lines in P, at latitude $\vartheta$ on the surface of Earth rotating with diurnal angular velocity $\vec\omega_{\oplus}$, have masses of different composition, $A$ and $B$. We assume them  to have   the same inertial mass $m_{i}$ but different gravitational masses: $m_{A}^{g}=m_{i}$, $m_{B}^{g}=m_{A}^{g}(1+\eta)$, with $\eta\neq0$ the \eotvos\ parameter  quantifying  the violation of equivalence. The figure shows their  deflections towards South (if $\eta>0$, then $\varepsilon_{A}>\varepsilon_{B}$). The figure is obviously not to scale. The deflection angle of a plumb line is very small: $ \varepsilon\simeq\frac{\omega_{\oplus}^{2}R_{\oplus}}{2g}\sin2\vartheta$, with  $R_{\oplus}$  the radius of Earth and $g$ the local gravitational acceleration, and  a maximum deflection of 
$\simeq1.7\cdot10^{-3}\,\rm rad$ at exactly $45^{\circ}$ latitude.
}
\label{Fig:DisegnoEotvos}
\end{center}
\end{figure}

 A breakthrough occurred in 1890~\cite{Eotvos1890} when \eotvos\ published the first results of his tests of the equivalence between inertial and gravitational mass to $5\times10^{-8}$ obtained  with test masses  suspended on a torsion balance rather than as simple pendulums. In 1909~\cite{EotvosPrize1909} he won a prize  of the University of G\"ottingen   reporting  an improved precision of $10^{-8}$ whose  details were published more than a decade later, after his death~\cite{Eotvos1922}. In the words of Einstein~\cite{Einstein1914Scientia}, who referred specifically to the fact that   \eotvos\ had  confirmed  the equivalence to  $10^{-8}$, the torsion balance experiment is described as follows:
\begin{quote}
 \textit{\eotvos' experimental method is based on the following. A body on the surface of the Earth is acted upon by the terrestrial gravitation and the centrifugal force. The gravitational mass is the determining factor for the first force, and the inertial mass for the second one. If the two did not coincide, then the direction of the resultant of the two (apparent gravitation) would depend on the material of which this body consists.}
\end{quote}

Fig.\,\ref{Fig:DisegnoEotvos}  is a graphical representation of Einstein's words.  If the two masses in the  figure, for which violation is assumed, are placed at the two ends of a uniform balance arm,  suspended at its center,  with radius vectors $\vec r_{A}$, $\vec r_{B}$ from the center of mass of the Earth ($\vec r=\vec r_{B}-\vec r_{A}$ is the balance arm), they are subjected to the  forces:  
\begin{equation}\label{Eq:ForcesViolation}
\vec F_{A}=m_{i}\vec g_{A}\ \ \ , \ \ \ \vec F_{B}=m_{i}\vec g_{B}
\end{equation}
which give  rise to the torque:
\begin{equation}\label{Eq:Torque}
\vec T=\vec r_{A}\times\vec F_{A}+\vec r_{B}\times\vec F_{B}\ \ \ .
\end{equation}

The total gravitational force from Earth $\vec F_{A}+\vec F_{B}$ is applied to the center of mass of the balance, and  is  counteracted by the tension of the suspension wire. Thus, at $1$-$g$ the unit vector along the direction of the wire is:
\begin{equation}\label{Eq:WireUnitvectorAt1g}
\hat w=-\frac{\vec F_{A}+\vec F_{B}}{|\vec F_{A}+\vec F_{B}|}                            
\end{equation}
and  the component of the torque $\vec T$ along $\hat w$,  which twists the balance until it is counteracted exactly by the restoring torque of the suspension fiber, is~\cite{TBfocusIssue2012}:
\begin{equation}\label{Eq:TorqueAlongWire}
T_{w}= \vec T\cdot\hat w=\frac{(\vec F_{A}\times\vec F_{B})\cdot\vec r}{|\vec F_{A}+\vec F_{B}|}   \ \ \ .                  
\end{equation}
Since $\vec F_{A}\times\vec F_{B}$ lies in the East-West direction of the horizontal plane, the effect of violation on the balance is maximum if its arm  is  in the same direction.

This formula shows that  only forces on the two masses which are not parallel to each other do twist the balance, while parallel forces --even if they have  different magnitude-- do not affect it.  
Thus, the torsion balance is an intrinsically differential instrument capable of rejecting common mode forces even of different magnitude  --as long as they are parallel.  Moreover, by using a very thin suspension fiber, its elastic torsional constant is very small (being inversely proportional to the $4$th power of the radius of the fiber), which makes the balance sensitive to extremely  tiny torques.

Test masses on the balance  are not concentric, therefore they are subjected to different forces because the gravitational field is not uniform (gravity gradient, or tidal effects). In the simple case of a balance with a single  arm and  one test mass at each end, even if the arm is exactly horizontal, i.e. perpendicular to the local vertical defined by the sum of gravitational and centrifugal force  (and assuming no violation),   the gravitational accelerations $\vec g_{A}$ and $\vec g_{B}$ on the two masses  are not parallel  to each other because the masses are separated by the non-zero relative vector $\vec r$.  Therefore, according to (\ref{Eq:TorqueAlongWire}),  there is a spurious tidal effect  which mimics a violation signal of order $r/R_{\oplus}\simeq2.3\cdot10^{-8}$  for a balance arm $r\simeq15\,\rm cm$.

However, if the balance arm is flipped, rotating it by $180^{\circ}$ around $\hat w$ (or else, the balance is kept fixed and the test masses are exchanged), the torque due to gravity gradient does not change sign. 
Instead, in case of violation, the torque $T_{w}$ caused by the forces (\ref{Eq:ForcesViolation}) would change sign, because  $\vec r\rightarrow -\vec r$, while the forces (\ref{Eq:ForcesViolation}) remain the same. As shown in Fig.\,\ref{Fig:DisegnoEotvos}, the mass which was deflected toward South more than the other because of its composition, will still be deflected more than the other after flipping. Thus, by inverting the masses (or by  flipping the balance) the deflection angle caused by a violation of equivalence changes sign while the one due to gravity gradient does not. 
By rotating the balance around an axis directed along $\hat w$ with frequency $\nu_{spin}$ (and $\vec r\bot\hat w$),  the violation signal from Earth appears at $\nu_{spin}$ while the tidal effect is at $2\nu_{spin}$, and they can be separated.
 \eotvos' balance was static and could only be flipped manually.  It is not surprising that he was able to reach about $10^{-8}$.

\section{From E\"OtV\"OS to E\"OT-Wash: up-converting the frequency of the signal by rotation}
\label{Sec:EotvosToEotwash}

Rotation of the torsion balance around the suspension fiber up-converts a violation signal  in the field of  Earth from zero (DC) to the rotation frequency, making it possible to detect a  deflection angle  at a known frequency and  to distinguish it from systematics. 
Moreover, it has long been known~\cite{Saulson1990} that in mechanical experiments  losses due to internal damping are lower at higher frequencies. Therefore, by up-convertion the signal finds itself in a region of reduced thermal noise, a well exploited property in experiments to test the weak equivalence principle~\cite{TBfocusIssue2012, PRLthermalnoise} (see Sec.\,\ref{Sec:StateOfArt} and Sec.\,\ref{Sec:GG}). 

Rotation of a very sensitive instrument such as the torsion balance has been regarded for a long time as likely to give rise to an unacceptable additional  noise. In the mid 1960s Dicke and collaborators~\cite{DickeEPtest1964} used  the Sun as source body of a possible violation, by comparing the gravitational and inertial mass of the test bodies  which enter, respectively,  in the gravitational force from the Sun and in the centrifugal force along the orbit around it.  The `passive' rotation of the balance  together with the Earth  at the diurnal frequency  makes a violation signal from the Sun to appear at this frequency, with no need to `actively' rotate the balance.

A disadvantage with the Sun as source is a slightly weaker driving signal ($ n^{2}d_{\oplus\odot}\simeq6\times10^{-3}\,\rm ms^{-2}$, with $n$ the annual angular orbital velocity of the Earth and $d_{\oplus\odot}$ the Earth-Sun distance) as compared to that in the field of the Earth ($\omega_{\oplus}^{2}R_{\oplus}/{2}\simeq1.7\times10^{-2}\,\rm ms^{-2}$). 
More importantly,   at the Earth   diurnal frequency at which a violation signal would appear, other disturbances are known to affect the torsion balance, such as  diurnal thermal effects, diurnal microseismicity, diurnal local mass motions. They are caused  by the Sun not through its gravitational force  but through its illumination and heating of the Earth's  surface and  atmosphere (possibly with some time lag), in the day/night, hot/cold cycle. 
Earth tidal effects are less relevant because they are  (to first order) DC. The main tidal effect from the Sun is  at twice the diurnal frequency. More importantly, solar tides are smaller than Earth gradients   because  the  Earth-Sun distance $d_{\oplus\odot}$ is much larger than the Earth's radius. For a balance arm of $15\,\rm cm$ the effect  of gravity gradients from the Sun results in a spurious `violation'    of  order  $ r/d_{\oplus\odot}\simeq10^{-12}$.  

Dicke's group reached a precision of $10^{-11}$, about three orders of magnitude better than \eotvos\ had done in the field of Earth.  A few years later Braginsky~ \cite{Braginsky,BraginskyManukin} in Moscow improved the test in the field of the Sun by another order of magnitude, to $10^{-12}$. 

The improvement achieved in Moscow was  possible for various reasons. The ability in suspending masses with tungsten fibers of very low torsional constant and very high mechanical quality made the balance extremely sensitive to tiny torques, and   thermal noise very low.   The balance carried $8$ test masses (4 in Al and 4 in Pt) at the vertices of a regular octagon (the Al masses on one side  and the Pt ones on the other). This configuration makes the balance sensitive only to the $5$th derivative of the gravitational potential, thus reducing the spurious torques resulting from the gravitational coupling between   the mass distribution of the torsion balance and the nonstationary local masses, having a prominent diurnal component for the reasons mentioned above.
Last but not  least, as Braginsky recalled during a visit by one of us (AMN), the  Institute of Physics  in Moscow was built on a very deeply rooted  rock so that its  basement could be used for experiments requiring a very quiet environment. The torsion balance was still in the basement of the Institute at the time of the visit, and records of   the balance oscillations impressed  on a uniformly rotating film (by means of a laser and a small mirror placed on the balance) were still visible in Braginsky's office. 

In 1986 \eotvos' tests in the field of  Earth became of interest again because of the possible existence of a new composition-dependent long range force which would act on distance scales accessible by these experiments~\cite{Fischbach1986}. Such a force could  not be ruled out by the much more precise tests in the field  of the Sun, hence pointing out the need to improve the old \eotvos' results. 

After Dicke's and Braginsky's experiments it was apparent that the challenge was  a rotating torsion balance, in order to achieve  in the field of the Earth the same precision  demonstrated in the field of the Sun.  The \eotwash\ group led by Eric Adelberger at the University of Washington in Seattle has successfully completed the task, finding no violation of equivalence to a precision of about $10^{-13}$ in the field of the Earth and to a few parts in $10^{13}$ in the field of the Sun with  experiments that  reached the level of thermal noise expected at the rotation frequency of the balance (see~\cite{TBfocusIssue2012} and references therein). 

The improvement over \eotvos' results in the field of Earth  is outstanding, by almost $5$ orders of magnitude.

 In addition to achieving an extremely smooth and quiet rotation (at about $1$ $\rm mHz)$, the authors faced the challenge of spurious torques caused by the gravitational coupling of the Earth and the nearby masses with the mass distribution of the suspended balance. 

The torques caused by such couplings may be too large when aiming at high level precision. More importantly, since external masses do not rotate with the balance, some couplings   appear at the rotation frequency and therefore compete  directly with the violation signal. For instance, the monopole of  Earth coupling with a non-zero quadrupole mass moment of the balance results in a torque with the same frequency and phase as the signal. 

In order to deal with this issue the suspended balance was carefully designed (and manufactured) in such a way as to minimize its most relevant mass moments. The  effects of external masses  were measured using  various different mass distributions of the balance. By amplifying on purpose, one at a time,  its  mass moments (starting from the most relevant ones), the measurement  provided  the torque resulting by its coupling with the external masses, which was then compensated by setting up an appropriate distribution of lead blocks in the vicinity of the balance around it. 
A similar technique has been used recently by Chinese scientists~\cite{CinesiTB2017}.
The procedure required re-checking and various iterations, and was a complex combination of numerical calculations and experimental measurements carried out with extreme care, until the balance was almost insensitive to mass couplings and sensitive only to a violation of  equivalence  from Earth or to a new composition-dependent force. Which in the end made possible an improvement by almost 5 orders of magnitude.

In the field of the Sun the improvement by the \eotwash\ group over Braginsky's experiments has been by less than 1 order of magnitude, to a few parts in $10^{13}$. Rotation of the balance moves the signal away from the diurnal frequency to a higher frequency at which thermal noise is lower. However, since the environment does not rotate with the balance, all local disturbances which are induced by the Sun at 1 day period, give rise to torques with the same frequency as the signal. 
As a result, the improvement over the results  obtained by exploiting only the `passive' rotation provided by the Earth  of a stationary balance relative to the Sun could not be as spectacular as in field of Earth.

\section{State of the art and limitations}
\label{Sec:StateOfArt}

At present the best tests of the weak equivalence principle in the gravitational field of  Earth  are those by the \eotwash\ group performed with rotating torsion balances to a precision of $10^{-13}$. In the field of the Sun the precision is slightly worse, of a few parts in $10^{13}$~\cite{TBfocusIssue2012}.

The gravitational effect of the Sun on bodies with different mass-energy content has been tested also with LLR experiments. Using  more than $45$ years of laser ranging data to retroreflectors on  the surface of the Moon scientists have shown no violation for  Earth and Moon  towards  the Sun to about $10^{-13}$\cite{LLRWilliams2012,LLRMueller2012}.  (The composition of the Moon is similar to that of the Earth's mantle and different from the Ni-Fe  rich core).

Rotating torsion balance experiments are limited by gravity gradients, particularly those changing with time (e.g. because of water flow, which is quite relevant in Seattle), and by thermal noise (losses in the suspension fiber at the rotation frequency of the balance). 

Further reduction of gravity gradient effects requires still more care in the balance design as well as  in the measurement and compensation of these effects. 

Thermal noise can be reduced by lowering the temperature, by increasing the rotation frequency or by manufacturing  a suspension fiber with lower internal losses. The  \eotwash\ group has tested a cryogenic balance, but its performance has never matched that of their room temperature balances. The current $\rm mHz$ rotation frequency appears to be the best choice for this mechanical oscillator in 1D (see~\cite{TBfocusIssue2012,PRLthermalnoise}). A fiber fabricated in fused silica, whose losses are expected to be much lower than in tungsten wires, is under consideration.  Should such lower  losses be confirmed also at the rotation frequency of the balance, and provided that  electric charging of the non-conductive fiber is sufficiently small, thermal noise will be reduced.

 In addition, the \eotwash\ group plans to use neutron-rich beryllium and proton-rich polyethylene test bodies because they  have a higher composition contrast which would result in a more precise WEP test for the same experiment sensitivity (see~\cite{Adelreview2009,GGphaseA22009,MikePRL2013}). 

 Altogether the \eotwash\ group is aiming at 1 order of magnitude improvement both in the Earth's and in the Sun's field~\cite{TBfocusIssue2012}.

LLR tests are affected by gravity gradients between the Earth and the Moon  in the field of the Sun. An error $\Delta {\rm a}$ in the measurement of the semimajor axis of the lunar orbit  yields a spurious `violation' signal of about $3\Delta{\rm a}/d_{\oplus\odot}$~\cite{EPwithlaserranging}. If laser ranging to the Moon has errors at  cm level, the limit from gravity gradients is of the order of $10^{-13}$. 

Considerable efforts have allowed laser ranging to the Moon to be improved by about a factor of $10$, with errors at mm level~\cite{APOLLOfocusIssue2012}, making it possible a similar improvement of  LLR tests of WEP. However, for this remarkable hardware achievement  to yield a comparable improvement   of  the WEP test  it is necessary that the physical model of the lunar orbit from which the effect of violation is obtained (a polarization of the orbit towards the Sun, absent if WEP holds) be improved too by 1 order of magnitude.

Satellite Laser Ranging (SLR) to Laser Geodynamics Satellites (LAGEOS)  orbiting the Earth at a distance of about two Earth radii  has been suggested as another possibility for testing WEP. In this case, because of the smaller distance from the source body (Earth) laser ranging errors  at cm level yield a larger spurious signal  of about $3\Delta {\rm a}/(2R_{\oplus})\simeq2.4\cdot10^{-9}$~\cite{EPwithlaserranging}. 

It is  the large distance from the Sun which allows LLR tests of WEP to reach $10^{-13}$, not only  the fact that in the Earth-Moon system `The heavy masses make it insensitive to any disturbances other than celestial perturbations',  as stated  in~\cite{BlaserReleaseErrors}. Non-gravitational forces are a matter of concern for artificial satellites because of their large area-to-mass ratio~\cite{MNF}, a parameter which is very small in the case of celestial bodies --such as the Earth and the Moon--  due to their large size.
In spite of that, the physical model needed to predict the motion of the Moon  on the basis  of laser ranging data from Earth-based stations to the laser reflectors on the lunar surface, is very complex and hard to improve.
From the hardware side, gravity gradient effects  at the distance of the Moon resulting from laser ranging errors   are  indeed the key limitation to WEP tests by LLR, and this is why scientists have worked very hard to reduce them from cm to mm level.

Similarly to the WEP test for the Earth-Moon system around the Sun, one might consider a similar test for the  Earth-LAGEOS system in the field of the Sun, since in this case the limitation by gravity gradients would be at $10^{-13}$ level, not $10^{-9}$. However,  this test would be  a factor 300 less sensitive than in the case of the Moon~\cite{EPwithlaserranging}. In essence, this is because LAGEOS is closer to the Earth than the Moon, therefore  its orbit is less affected  than the lunar orbit by the Sun, which is also the source mass of a possible violation.
The same holds for the more recent LARES satellite,orbiting at lower altitude than  LAGEOS.

LLR tests of WEP are therefore not going to be  superseded  by similar SLR tests, and expectations are for one order of magnitude improvement.

Since Earth and  Moon have different non-negligible self-gravitational  binding energies LLR can test   the property of gravity itself to obey WEP, a test which is obviously beyond the reach of all experiments with artificial test bodies.  GR requires self gravitation to obey WEP,  while other metric theories of gravity do not, hence these tests can discriminate. Since self gravitation is  a  very small fraction of the total mass-energy even for celestial bodies, tests of whether it obeys  WEP  are much less precise. LLR has reached a few parts in $10^{4}$\cite{LLRMueller2012}. Tests of UFF for gravity are performed also in strong field with pulsar-white dwarfs binary~\cite{Freire2012} and recently with the triple pulsar data for which a result to $3\cdot10^{-8}$ is reported~\cite{TriplePulsar2016}.  

Although WEP tests to the highest possible precision are the deepest probe of GR and  the existence of a composition-dependent new force, testing WEP for various forms of mass-energy and as many different compositions as possible, even with low precision, is very important. 

With  the explosion of supernova 1987A in the Large Magellanic Cloud, observation of a neutrino burst within a few hours of the associated optical burst  provided a test of the weak equivalence principle. The observed delay was used to conclude that different particles (neutrinos and photons) undergo the same effect (the Shapiro time delay) from the gravitational field of our galaxy to a few parts in   $10^{3}$~\cite{NeutrinosPhotons1,NeutrinosPhotons2}. 

At CERN scientists of the AE$\bar g$IS collaboration  plan to measure the local gravitational acceleration of antihydrogen to $1\%$, thus performing for the first time a direct test of the weak equivalence principle with antimatter~\cite{AEGISbern}.

Another class of WEP tests  are the so-called mass drop tests, which have not been used until the late 1980s. Galileo himself, despite the legend,  tested UFF with pendulums and not by dropping masses from a height. In 1986 the need emerged to improve \eotvos' tests of WEP in the field of  Earth~\cite{Fischbach1986}  and various groups attempted to do so by dropping masses instead of using a torsion balance. It was clear from Dicke's and Braginsky's experiments that rotation was the key to improve \eotvos' results  in the field of Earth,  but rotating a very sensitive balance in the lab  was regarded as too challenging. On the other hand, the much larger driving acceleration,  by almost 600 times ($g\simeq9.8\,\rm ms^{-2}$ in mass drop  tests versus $g\varepsilon\lesssim1.69\cdot10^{-2}\,\rm ms^{-2}$  on the torsion balance), and  the availability of high resolution, low noise readout based on  laser interferometry made mass drop tests very attractive. Non-zero gravity gradients due to initial condition errors (test masses starting from different heights and with different velocities) were a known competing effect  which scientists  hoped to minimize by  devising clever arrangements of the test masses.

In spite of all efforts, and regardless of a factor of about 600  in their favour,  drop tests achieved precisions in $\Delta g/g$ of several parts in $10^{10}$~\cite{NiebauerFaller,Kuroda,ErseoGAL}. They  would not compete with the \eotwash\ rotating balance and even the improvement over the torsion balance tests of \eotvos\  was quite modest, a factor of about 14 despite the sophisticated  technologies employed. In terms of sensitivity to differential accelerations, the \eotvos's  balance was still more sensitive (its driving signal being about 600 times weaker).  
As discussed by~\cite{ErseoGAL,NobiliBern} the limitation was due to initial condition errors at release which, because of  the gravity gradient of the Earth,  give a differential acceleration error which mimics a violation signal.

A longer free fall time would help, since the effect of a violation increases quadratically with time. Drop tests  using balloons and sounding rockets have been proposed~\cite{Iafolla, POEM}. In~\cite{POEM} the plan was to make drops with reversed axis in order to distinguish the effect of gravity gradient.  In~\cite{Iafolla} the test masses are coupled to form a sensitive differential accelerometer rotating around a horizontal axis  to be dropped inside a capsule carried by a balloon with $ 30\,\rm s $ free fall time. In this case rotation would make gravity gradient to appear at twice the rotation frequency while a violation signal would be at the rotation frequency, so that they can be separated.

Drop tests with cold atoms have  recently  reached  $10^{-8}$~\cite{WEPcoldAtomsChinese2016}.  These experiments use known techniques of quantum mechanics to test  the universality of free fall with  two atom clouds   made of  atoms of  different species (or just different isotopes). What is being tested is whether different atom species (no matter, in principle, how many are contained in each cloud) couple in the same way, or not, with the gravitational field of Earth. As mass drop tests they face the same problems as drop tests with bulk masses, though the way such problems can be solved must take into account the actual features of the experiment.

At the $10^{-8}$ level the systematic effect of relative initial condition errors in the presence of the gravity gradient of Earth does not matter, but it does when aiming at higher precision.

A cold-atom drop test is underway inside a 10-m-tall vacuum chamber, with a  longer free fall time than other ground tests~\cite{Kasevich2007,Kasevich2013}. The precision of the test will depend also on the ability to deal with the gravity gradient/initial condition issue. 

\section{From ground to space}
\label{Sec:GroundToSpace}

More than 400 years since Galileo's pendulum experiments in Pisa at the start of 1600, the equivalence of inertial and gravitational mass is confirmed to a precision 10 orders of magnitude better, to $10^{-13}$. As shown in Sec.\,\ref{Sec:StateOfArt}, from now on we can  expect  at most a 10-fold improvement.

The torsion balance is extremely sensitive and very effective in rejecting common mode forces; being manufactured with  tolerances at the $10^{-5}$ level~\cite{TBfocusIssue2012} its  measurements of differential effects reach a relative precision of $10^{-13}$. The problem  is that the driving signal relative to which the balance  measures extremely  small differences, is  small. As shown in Fig.\,\ref{Fig:DisegnoEotvos}, the gravitational and inertial mass under scrutiny  involve only the small component $g\varepsilon$ of the  gravitational acceleration in the horizontal plane arising to counteract, with a tiny deflection $\varepsilon$,  the horizontal component of the centrifugal acceleration  due to the diurnal rotation of Earth. 

Instead, for a mass in orbit around the Earth the centrifugal force (proportional to its inertial mass)  equals in magnitude the total gravitational force from Earth (proportional to its gravitational mass), which at low altitude is not much lower than on ground.  

An instrument carried by a dedicated spacecraft in orbit around the Earth at altitude $h$, and  capable of  detecting a differential acceleration $\Delta a$ between two masses of different composition,  will test WEP to $\eta=\frac{\Delta a}{g(h)}$, with $g(h)=\frac{GM_{\oplus}}{(R_{\oplus}+h)^{2}}$. With the same sensitivity as the \eotwash\ balance and $h\simeq630\,\rm km$ it would test WEP almost 500 times better,  because the driving signal in orbit $g(h)$ is almost 500 times stronger than $g\varepsilon\lesssim1.69\cdot10^{-2}\,\rm ms^{-2}$ on ground. A further improvement in sensitivity to differential accelerations by a factor of  $20$  would yield a WEP test in the field of Earth to $10^{-17}$. With the Sun as source the improvement would come only from a better sensitivity to differential accelerations, because in this case  the driving signal in low Earth orbit is not stronger than on ground.

In principle a WEP experiment in orbit  has considerable advantages. At  (almost) zero-$g$ masses can be suspended  with a very weak coupling constant, which means  high sensitivity. As torsion balances have shown, rotation is a key  feature of WEP experiments.
Unlike on ground, in space the entire laboratory (the spacecraft) is an (almost) isolated system, hence local `terrain' tilts and microseismicity are much reduced. In addition, the lab co-rotates with the instrument (which would be impossible on ground) needing no stator and no bearings (which eliminates a major source of noise typical of rotating experiments on ground);
 if the spacecraft is one-axis stabilized,  as it is the case  with GG,  rotation occurs by angular momentum conservation after initial spin up, and not even thrusters are needed (`passive' rotation, similarly to Earth's diurnal rotation). Co-rotation also eliminates altogether the effects of local mass anomalies (as long as there are no moving parts) because they are DC. 
 
 A major issue to deal with in space, which is not present on ground, are non-gravitational forces acting on the outer surface of the free falling spacecraft, such as solar radiation pressure and drag due to residual air along the orbit, which dominates at low altitudes. For WEP experiments in space this is not a small perturbation, but an effect many orders of magnitude larger than the target signal. Microscope deals  with it  by drag-free control (see Sec.\,\ref{Sec:Microscope}); in order to reach higher precision, GG exploits, in addition to drag-free control, the possibility to reject common mode effects (such as the inertial forces resulting from drag)  by designing the test masses as a variant in space of the torsion balance (see Sec.\,\ref{Sec:GG}).

How would a torsion balance work in space?

A practical problem is that in absence of weight it needs to be suspended from both ends. On ground the fiber defines the local vertical for a balance with a given distribution of  different composition test masses; its direction is therefore defined  by (\ref{Eq:WireUnitvectorAt1g}) leading to the fact, expressed by  (\ref{Eq:TorqueAlongWire}), that only forces with different directions twist the balance.  For the torsion balance the  property of   high common mode rejection is thus provided mostly by physics.  At zero-$g$  this property is lost, and one must rely on precision manufacturing,  possibly with some adjustments in flight.  In space, with the balance on an orbit that cannot be exactly circular,  gravity gradient effects at the same frequency as the signal  on test masses  a few cm apart would  be far too large to be acceptable.  

The only way to deal with gravity gradient (tidal) disturbances in space is to have (nominally) concentric masses, in practice two cylinders --of different composition-- one inside the other (see Sec.\,\ref{Sec:Microscope} and Sec.\,\ref{Sec:GG}).

On ground drop tests have been unable to compete with the rotating torsion balance by far. In space they have  no  factor to gain in strength of the driving signal, which is in fact slightly weaker than it is for drop tests on ground. Some gain would come from a longer time of fall in absence of weight. However,  Earth  tidal  effects due to systematic initial condition errors which limit drop tests on ground,  are  an issue to be faced also in space. Nonetheless, cold atoms drop tests in space have been investigated~\cite{STEQUEST,STEQUEST2014} with the goal of improving the results of similar ground experiments~\cite{OneraWEP2013,RaselEP2014},~\cite{WEPcoldAtomsChinese2016}  by 50 million and 5 million times respectively, to reach $2\cdot10^{-15}$. 
At this level the small number of atoms in the  clouds compared to Avogadro's number, combined with the gravity gradient/initial conditions errors issue, result in  a fundamental limitation~\cite{ColdAtoms2016}.  A possible way out  is to reduce and/or  separate the effect of the gravity gradient of Earth.  To this end, an attempt at using methods similar to those employed in experiments with bulk masses has recently been proposed~\cite{PRAgradients2017}. However, substantial differences between the two kinds of experiments show that the proposed solution  would not be as effective as hoped~\cite{AAgradienti}.

\section{The Microscope experiment in orbit aiming at $10^{-15}$}
\label{Sec:Microscope}


The Microscope satellite was launched on 25 April 2016 with the goal of testing the weak equivalence principle in the field of Earth  to $1$ part in $10^{15}$~\cite{MicroscopeFocusIssue2012}. It has successfully completed the commissioning phase  and since December 2016  is taking scientific data~\cite{OneraPressRelease161207}.  Preliminary results have been reported at the 2017 conference on Gravitation in La Thuile, Italy.

\begin{figure}[h!]
\begin{center}
\includegraphics[width=0.30\textwidth]{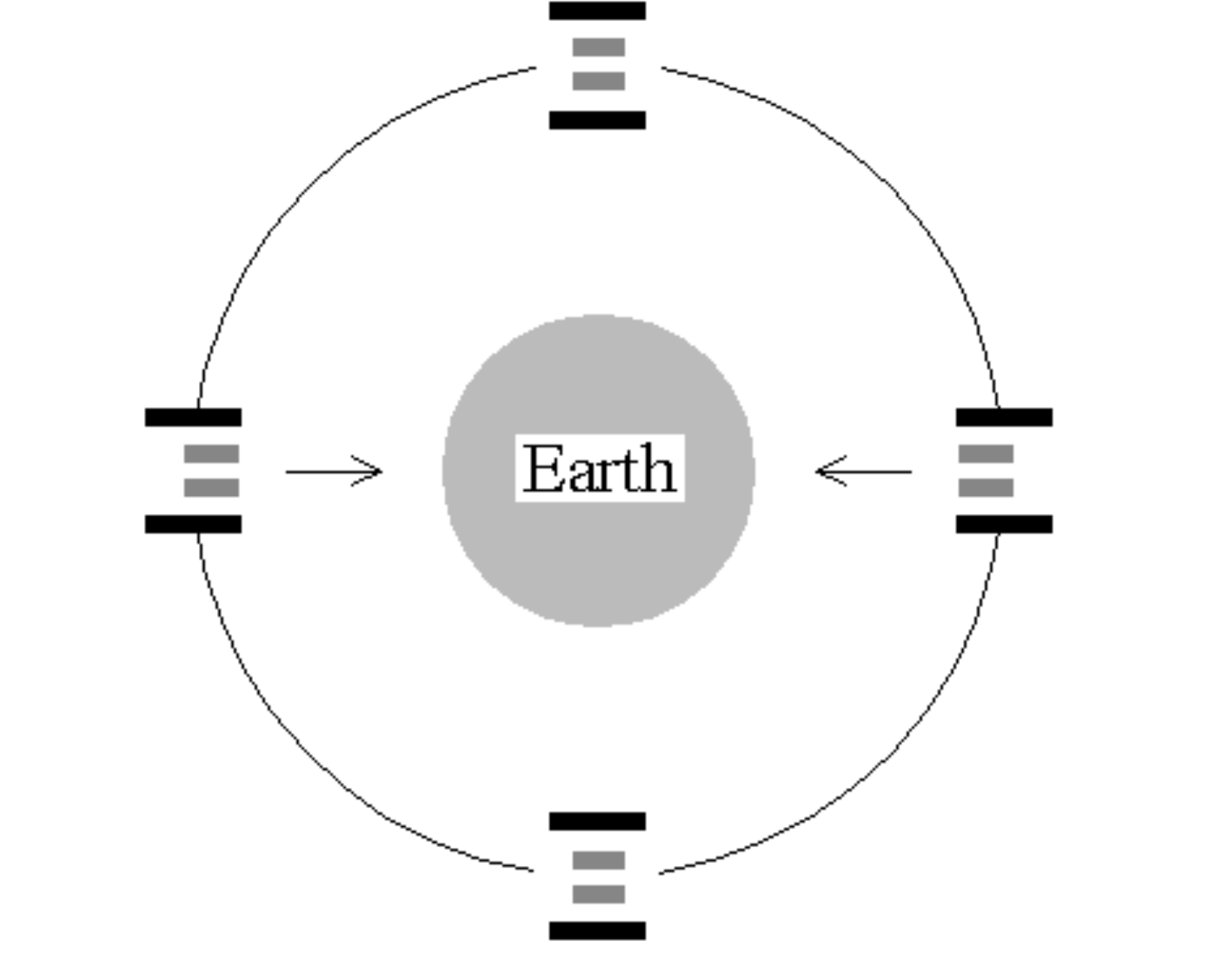}
\caption{Principle of Microscope test of the weak equivalence principle. 
Two concentric, co-axial test cylinders of different composition move on a circular Earth orbit with the symmetry (and sensitive) axis lying in the orbit plane, inside a spacecraft (not shown) whose attitude is actively kept fixed relative to inertial space. It is the original concept proposed at Stanford~\cite{WordenSTEP1973,WordenSTEPthesis1976,Worden1978}, but in Microscope the cylinders are at room temperature rather than  cryogenic at superfluid He temperature.
Electrostatic forces suspend each cylinder individually  in such a way that it is allowed to move  only along the symmetry axis, the electrostatic stiffness being low along the axis and high in the plane perpendicular to it. For each cylinder the displacement relative to the enclosure (rigid with the spacecraft, not shown either) unbalances the capacitance bridge and allows the displacement to be measured~\cite{Onera1999}
The sketch shows  a violation of equivalence whereby the inner test cylinder  is affected by a slightly stronger acceleration than the outer one towards the  center of mass of Earth. The violation signal is at the orbital frequency $\nu_{orb}\simeq1.7\cdot10^{-4}\,\rm Hz$. (Figure not to scale; the satellite orbits at $h\simeq700\,\rm km$ altitude).
}
\label{Fig:MicroscopeViolationSignal}
\end{center}
\end{figure}

The principle of Microscope experiment is shown in Fig.\,\ref{Fig:MicroscopeViolationSignal}. It was originally proposed by Paul Worden at Stanford in the 1970s~\cite{WordenSTEP1973,WordenSTEPthesis1976,Worden1978}  for STEP, a cryogenic space  test of the WEP to be carried out at superfluid He temperature  to a precision of 1 part in $10^{17}$.

With a driving acceleration at the Microscope altitude $g(h)\simeq8\,\rm ms^{-2}$,   a WEP test to $10^{-15}$  requires to detect a differential acceleration between the test cylinders $\Delta a_{_{WEP}}\simeq8\cdot10^{-15}\,\rm ms^{-2}$. The corresponding relative displacement depends on the stiffness of the suspension.
As reported by~\cite{Rodrigues2009} (see also~\cite{Touboul2009}), the frequency induced by electrostatic stiffness along the sensitive axis is  $1.45\cdot10^{-3}\,\rm Hz$ for the outer cylinder and $9.76\cdot10^{-4}\,\rm Hz$ for the inner one. As a result, a violation of equivalence to $10^{-15}$ would displace one test cylinder relative to the other by about $100\,\rm pm$. 

\begin{figure}[h!]
\begin{center}
\includegraphics[width=0.26\textwidth]{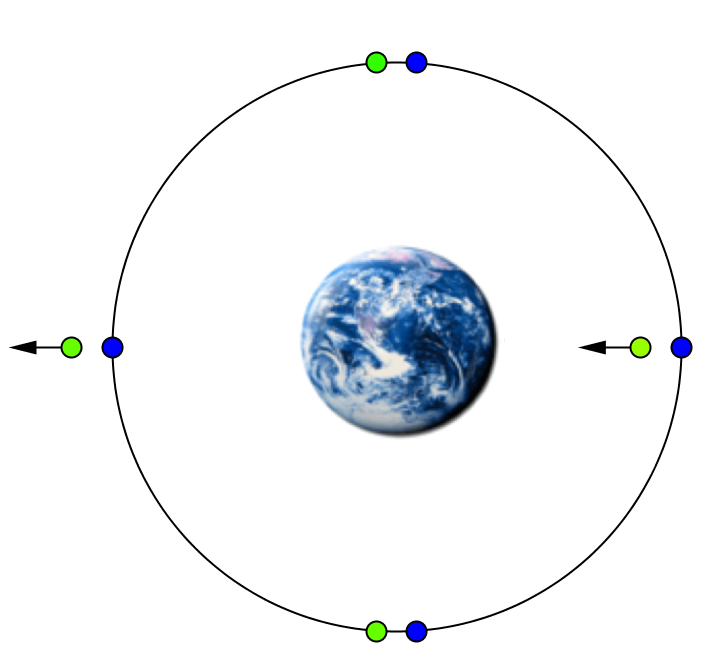}
\caption{Sketch of the effect of  the gravity gradient of Earth if the centers of mass of the inner and outer test  cylinders shown in  Fig.\,\ref{Fig:MicroscopeViolationSignal} (depicted here as a green and a blue dot) are   not exactly coincident  along the sensitive axis because of  construction errors. The arrows represent the dominant  tidal acceleration  of the green test cylinder relative to the blue one due to  its different radial distance from Earth. If the spacecraft attitude is fixed in space and the orbit is perfectly circular, this acceleration is the same every  half orbital period, i.e. its frequency is $2\nu_{orb}$ while the violation signal would be at $\nu_{orb}$, as shown in Fig.\,\ref{Fig:MicroscopeViolationSignal}.}
\label{Fig:MicroscopeGravityGradientsTides}
\end{center}
\end{figure}

Because of construction errors --mostly due to the complex assembly of its electrodes-- each cylinder is not exactly centered along the sensitive axis  and there is an offset between the two, giving rise to tidal effects from Earth. As shown in Fig.\,\ref{Fig:MicroscopeGravityGradientsTides}, the largest tidal effect is at twice the orbital (and signal) frequency, amounting to $\Delta a_{tide}\simeq\frac{2GM_{\oplus}}{R_{\oplus}^{3}}\Delta x_{off} $. Microscope requires  $\Delta x_{off}\simeq20\,\mu\rm m$ ~\cite{MicroscopeFocusIssue2012}, in which case it is $\Delta a_{tide} \simeq4.5\cdot10^{-11}\,\rm ms^{-2}$. 
The offset measured in orbit is only slightly larger than required~\cite{OneraPressRelease160927}.

With the electrostatic stiffness reported above, the corresponding tidal displacement is  $\Delta x_{tide}\simeq0.6\,\mu\rm m$. If the orbit is not perfectly circular, a fraction of the main  tidal effect proportional to the eccentricity will be at the orbital frequency, hence competing with the signal. Luckily, the eccentricity of Microscope's orbit has turned out to be lower than required,  more than compensating for the larger offset between the centers of mass of the test cylinders.

The way how Microscope  deals with the tidal effect at the signal frequency is, similarly to STEP, to use the main tidal effect at $2\nu_{orb}$  predicted  by celestial mechanics  to estimate in flight the offset which has generated it to $1/200$ of its value ($\simeq0.1\,\mu\rm m$)  and then \textit{a posteriori}, during data analysis, separate  from the signal  the tidal effect  at the same frequency (see~\cite{MicroscopeFocusIssue2012} for the physical parameters whose knowledge is needed for this procedure to work).  Although gravity gradients inside a small lab isolated in space away from Earth (the spacecraft) are not so much a concern as they are for torsion balances on ground,  they are still a key issue to deal with when aiming at a high precision test of the WEP.

WEP experiments  in  space are affected by non-gravitational forces acting on the outer surface of the spacecraft and not  on the test cylinders weakly suspended  inside it. In low Earth orbit the largest one  is due to residual air drag, giving rise to an inertial acceleration on each test cylinder equal and opposite to the  air drag acceleration of the spacecraft; the largest  drag  component  is at the orbital frequency --like the signal-- and many orders of magnitude larger. 

Although in principle each test mass  should acquire the same inertial acceleration, individually suspended cylinders  as in Microscope are very hard to match, thus leaving  large differential residuals (poor common mode rejection). In such a case, almost the entire effect of air drag  must be compensated by means of thrusters --whose disturbances must be taken care of-- which force the spacecraft to follow  the common mode motion of the two test cylinders (drag-free control).

Each  cylinder sensitive along the symmetry axis is subjected also to a direct non-gravitational acceleration known as `radiometer effect'.  A non-zero residual pressure inside the enclosure, combined with a non-zero temperature gradient along the axis originated by radiation from Earth, results in a spurious acceleration at the orbital frequency, like the signal~\cite{RadiometerPRD2001,RadiometerMicroscope}.

In Microscope  each cylinder is actively controlled using as actuators  capacitors similar to those which  detect its motion, otherwise it will hit the enclosure  and therefore end the experiment. 

The electrostatic  forces which control  the two cylinders so that  they are forced to follow  the same orbit, ultimately yield the accelerations of the two cylinders relative to each  other, including the one due to a violation of equivalence, if present. The control force  of each  cylinder contains (primarily):  a DC term, known as back action~\cite{Onera1999},  due to  the misbalance of the capacitors because of construction errors; a component at $2\nu_{orb}$ necessary to counteract the main gravitational  tidal effect; a component  at $\nu_{orb}$  which counteracts the residual inertial acceleration after drag-free control, the tidal effect due to a non-zero orbital eccentricity, the violation signal itself and also --if large enough to compete-- the non-gravitational radiometer acceleration. 

As reported in~\cite{Rodrigues2009}, the back action acceleration has been estimated  by Microscope scientists to amount to $1.85\cdot10^{-9}\,\rm ms^{-2}$  for the outer cylinder (with a misbalance error of $22.3\,\mu\rm m$)  and to $7.64\cdot10^{-10}\,\rm ms^{-2}$ for the inner one (with an error of $20.3\,\mu\rm m$).  Its fluctuations have a component at the signal frequency that must be taken into account.

Fig.\,\ref{Fig:MicroscopeViolationSignal} shows that by maintaining the spacecraft attitude fixed relative to inertial space a WEP violation signal appears at the orbital frequency $\nu_{orb}$. If the spacecraft rotates relative to inertial space at frequency $\nu_{spin}$  around an axis perpendicular to the orbit plane, the sensitive axis of the accelerometer rotates with it and the signal appears at the sum (or difference) of the two frequencies. Being perpendicular to the symmetry axis of the test cylinders, the rotation axis is unstable for small perturbations, hence rotation must be ensured actively, and it  is slow in order to add as little noise as possible. 

Microscope was originally planned to be operated partly in inertial mode (no rotation) and partly in rotation mode, with a spin frequency up to a few times faster than the orbital one (and in opposite direction) so as to up-convert the signal to  $\nu_{_{WEP}}=\nu_{orb}+\nu_{spin}$  reaching  almost $10^{-3}\,\rm Hz$\,[6], a value which is close to the rotation frequency of the \eotwash\  torsion balance.

 As reported recently, it has been found that a spin rate a few times faster than the maximum planned (by a factor 3.9) results in a better sensitivity. Thus, the faster rotation rate has been adopted, resulting in a signal frequency at $3.1\cdot10^{-3}\,\rm Hz$. 
Despite the higher amount of propellant required (and consequent shorter duration of the mission) Microscope scientists have concluded  that overall `the balance is in favor of spinning faster'. The inertial mode is no longer mentioned and preliminary results at the faster spin rate shall be announced soon.

Though this result came as a surprise to people who regard rapid rotation as incompatible with  precision experiments,  in fact it is not. Unlike on ground, rotation in space needs neither stator nor bearings because the whole spacecraft spins together. Thus, even in the case in which rotation requires thrusters and propellant because  the spacecraft  is not stabilized by conservation of angular momentum around a stable axis, nonetheless, the absence of stator and bearings makes rotation much less noisy than in any rotating experiment on ground.

 Furthermore, a faster  rotation helps in reducing thermal disturbances, which is  what Microscope scientists also report, including  a smaller radiometer  effect.

 Other effects such as gravity gradients, residual drag, radiometer etc., are also up-converted along with the signal. Thermal noise due to internal damping is lower at higher frequencies~\cite{Saulson1990}. In Microscope it is dominated by losses in the thin gold wire connecting each cylinder to the enclosure, with a quality factor of about $100$ at the frequency of interest~\cite{Touboul2009}. For a signal to noise ratio SNR=2 the integration time to reach the target of $10^{-15}$ is about $1.4\,\rm d$ ($20$ orbits). 
 
 Microscope carries two pairs of test cylinders. In the second  pair the two cylinders are made of the same material. An effect with the known frequency and phase of the signal which were detected by the different composition accelerometer (as shown in Fig.\,\ref{Fig:MicroscopeViolationSignal}) and not by the equal composition one, would be a violation signal. 
 This is correct in principle, but other differences between the two accelerometers (such as the fact that they are several cm apart in the nonuniform gravitational field of Earth, or a different  response to non-gravitational forces such as the radiometer effect~\cite{RadiometerMicroscope}) might yield somewhat different results even in absence of a violation. The equal composition accelerometer is an additional tool  to be used  together with   systematic error checks in order to distinguish spurious effects from a  violation signal.

\section{The `Galileo Galilei' (GG) candidate space experiment to test the WEP to  $10^{-17}$}
\label{Sec:GG}
`Galileo Galilei' (GG), to be flown in  low Earth orbit as Microscope,  aims at testing the weak equivalence principle to $1$ part in $10^{17}$ at room temperature~\cite{GGfocusIssue2012}.

GG can target a precision   100 times better  than Microscope because of its design. Table\,\ref{Tab:GGvsMicroscope} shows in summary a quantitative comparison between the two experiments. Two are the key features which make GG different from Microscope:  $i)$ it spins $1320$ times faster ($340 $ times faster  after the recently adopted higher spin rate), up-converting  the signal from the orbital frequency  to $\nu_{spin}\simeq1\,\rm Hz$ ($\nu_{spin}\gg\nu_{orb}$)  where thermal noise is much lower; $ii)$ the test cylinders ($10\,\rm kg$ each, for lower thermal noise and reduced non-gravitational  effects) are arranged to form a balance whose arms can be adjusted in flight in order to reject  common mode effects.

\begin{figure}[h!]
\begin{center}
\includegraphics[width=0.25\textwidth]{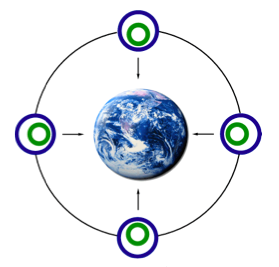}
\caption{Principle of the GG test of the weak equivalence principle. 
Two concentric, co-axial test cylinders of different composition move on a circular Earth orbit with the symmetry axis perpendicular to the  orbit plane, inside a spacecraft (not shown) whose attitude is stabilized by passive $1$-axis rotation at $\nu_{spin}\simeq1\,\rm Hz$ around the symmetry axis. 
%
The cylinders form  a  mechanical oscillator sensitive to differential forces in the plane  perpendicular to it,  and  much stiffer along the axis (Fig.\,\ref{Fig:GGbalance2DSection}).
The sketch shows  a violation of equivalence whereby the inner  (green) cylinder  is affected by a slightly stronger acceleration  than the outer (blue) one. The violation signal is a vector of constant size (for a perfectly circular orbit)  pointing to the Earth's center of mass while orbiting around it at  $\nu_{orb}\simeq1.7\cdot10^{-4}\,\rm Hz$. 
The experiment is designed in the framework of Newtonian dynamics because GR effects taking into account rotation of both the  test cylinders and the  Earth have been shown to be negligible by far~\cite{GRspinEffects2016}.
(Figure not to scale; the satellite orbits at $h\simeq630\,\rm km$ altitude).
}
\label{Fig:GGsignal}
\end{center}
\end{figure}
\vspace{-0.5cm}

\begin{figure}[h!]
\begin{center}
\includegraphics[width=0.41\textwidth]{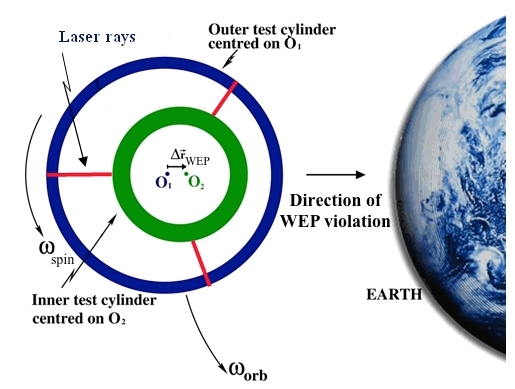}  
\caption{
The sketch shows a relative displacement vector $\Delta\vec r_{_{WEP}}$ between the centers of mass of the two cylinders $O_{1}$ and $O_{2}$ caused by  a violating differential acceleration driven by Earth $\Delta\vec a_{_{WEP}}$ ($\Delta\vec r_{_{WEP}}\simeq\frac{\Delta\vec a_{_{WEP}}}{4\pi^{2}\omega_{dm}^{2}}$).  
 The red lines depict the laser rays (at $120^{\circ}$ from each other) of the  interferometry gauge designed to read the relative displacements of the test cylinders. The whole system co-rotates at   $\nu_{spin}\simeq1\,\rm Hz$, therefore the signal  $\Delta\vec r_{_{WEP}}$  is read at $\nu_{spin}-\nu_{orb}\simeq\nu_{spin}$.
The cylinders are weakly coupled  as a 2D harmonic oscillator with normal mode differential frequency $\omega_{dm}$, hence equilibrium is reached  at the displaced position  $\Delta\vec r_{_{WEP}}$. This is the physical observable measured by the laser gauge, yielding the  \eotvos\ parameter $\eta=\frac{\Delta a_{_{WEP}}}{g(h)}$ that quantifies the level of WEP violation.  For the GG target  $\eta=10^{-17}$ the displacement signal to be measured is $0.6\,\rm pm$. 
%
 %
}
\label{Fig:GGsignalAndUpconversion}
\end{center}
\end{figure}

Rapid spin is possible by turning the symmetry axis of the test cylinders by $90^{\circ}$ with respect to Microscope (compare Fig.\,\ref{Fig:GGsignal} with Fig.\,\ref{Fig:MicroscopeViolationSignal}), making it  the (stable) rotation axis of the whole spacecraft which  stabilizes its attitude,  while the plane perpendicular to it is the sensitive plane of the test cylinders. The spacecraft conforms to the cylindrical symmetry imposed by the test bodies  and the whole system co-rotates passively (by conservation of angular momentum after initial spin-up) up-converting the signal from the orbital to the spin frequency by a factor of about $5800$ (see Fig.\,\ref{Fig:GGsignal} and Fig.\,\ref{Fig:GGsignalAndUpconversion} ).

\vspace{-0.3cm}

\begin{figure}[ht]
\begin{center}
\includegraphics[width=0.51\textwidth]{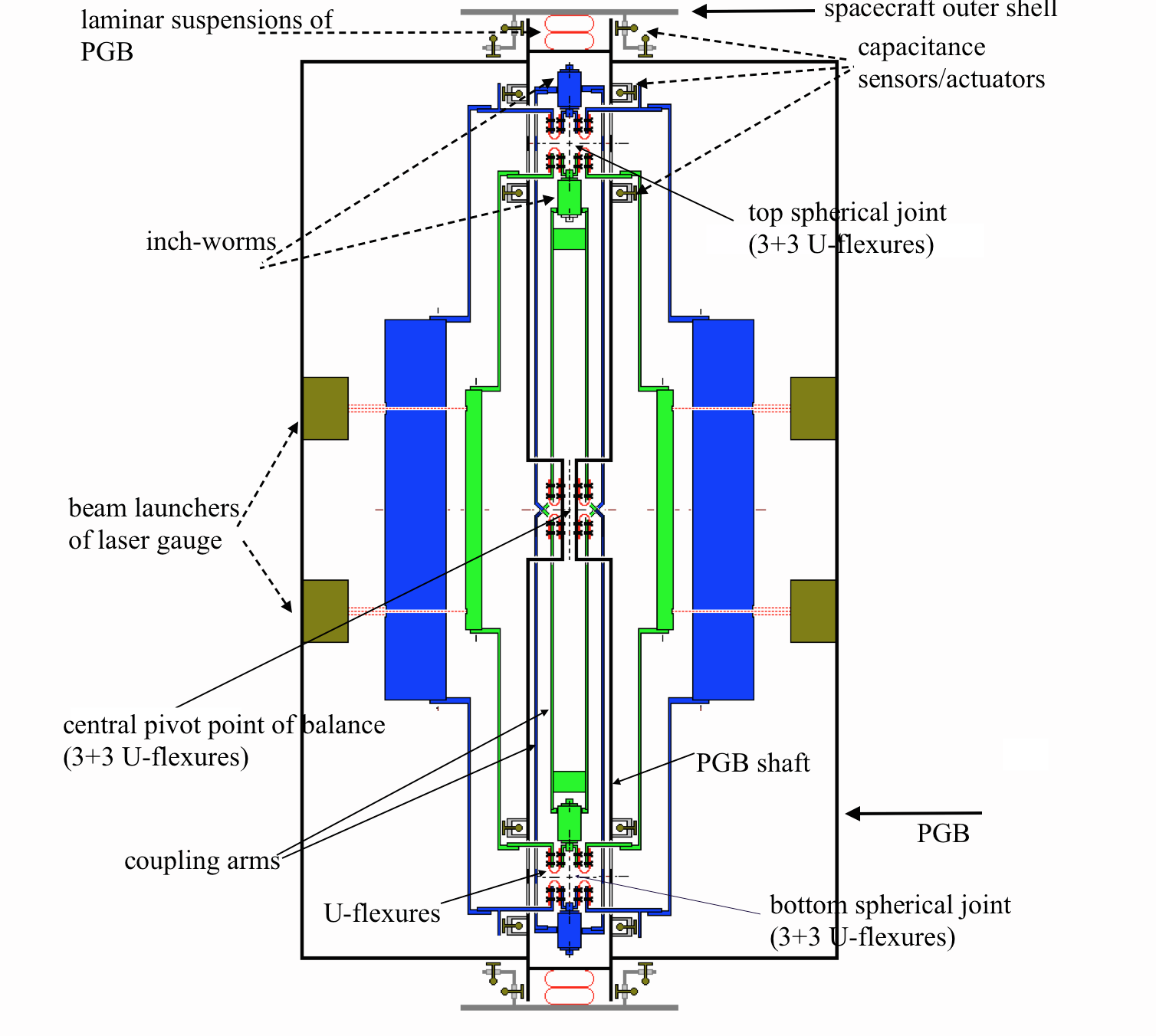}
\caption{Sketch of the GG balance made of 2 coaxial concentric test cylinders of different composition (in green and blue) spinning around the symmetry axis and weakly coupled in the plane perpendicular to it: section along the spin axis. An animation of this balance, oscillating with $1\,\rm s$ spin period under the effect of a  violation signal  is available on the GG webpage~\cite{GGwebpage}. In the animation the  oscillation amplitude is largely exaggerated: a violation to $10^{-17}$  would cause an oscillation of $0.6\,\rm pm$, which is only  $3\times10^{-11}$ the size of the $2\,\rm cm$ gap between the outer surface of the inner cylinder and the inner surface of the outer one.
The s/c is passively stabilized by one-axis rotation at $\nu_{spin}\simeq1\,\rm Hz$, nutation damping being provided by the weak coupling between the spacecraft and an intermediate stage (the PGB-Pico Gravity Box, and its shaft)  to which the balance is connected at the center by means of U-flexures (in red). 
This is the pivot of the balance. 
The motion of PGB relative to the s/c is read by capacitance sensors and drives the drag-free control system. 
At its two ends each test cylinder   is connected  (through a light rigid interface and 3 U-flexures $120^{\rm o}$  apart; 2 shown in planar section)  to the 2 ends of the coupling arms; the short ones for the inner green cylinder; the long ones for the outer blue cylinder.  The result are two  spherical joints (top and bottom) such that  any differential force acting between the test cylinders in the plane perpendicular to the symmetry axis will displace their centers of mass by tilting the coupling arms (pivoted at the center).  %
The laser gauge  boxes are fixed on the PGB and shown in brown. At  each end of the coupling arms (in blue and green) are shown the inch-worms which allow the balance to be balanced in order to reject accelerations acting in common mode on both cylinders.  The  two shorter parts of each coupling arm (pertaining to the inner cylinder and shown in green) have a small additional mass each (in green) so that the pivot center is at their center of mass. In the balance the mass of the test cylinders dominates over the mass of the coupling arms and interfaces. Note the  symmetry of the balance both in azimuth and top/down. 
This clever design of a beam balance with concentric test masses and perfect symmetry, as needed in space for testing the weak equivalence principle, is due to Donato Bramanti.}
\label{Fig:GGbalance2DSection}
\end{center}
\end{figure}

The way how two concentric  cylinders are coupled to form  a  beam balance is shown in Fig.\,\ref{Fig:GGbalance2DSection}.
  By balancing its arms in flight  against the common mode effect of drag  using  piezo actuators, the balance allows all common  mode effects (primarily the effect of drag) to be rejected. Since this is $50$ million times weaker than local gravity for ordinary balances on ground, it is apparent that balancing in space is much easier than on ground, where it has been known since a long time that balances can detect  differences of weight by $10^{-7}-10^{-8}$\,\cite{Quinn1993}.
Thus, unlike in Microscope where drag can only be compensated actively by the drag-free control system, in GG drag is partially compensated  and partially rejected by the balance, leaving a residual differential effect half the size of the target signal of GG (Table\,\ref{Tab:GGvsMicroscope}, entry III). 

The issue of a balance in space,  and the original GG setup, were the subject of lively discussions during a few days visit to Professor Braginsky in Moscow by one of us (AMN). He believed that a high precision test of WEP needs  a balance, in space as well as on ground, and strongly advocated that the test masses  be kept as passive and undisturbed as possible  (Table\,\ref{Tab:GGvsMicroscope}, entry X).

Thermal noise due to internal damping decreases as the square root of the  frequency~\cite{Saulson1990}, as confirmed by rotating torsion balances~\cite{TBfocusIssue2012}. The GG balance of Fig.\,\ref{Fig:GGbalance2DSection} is in essence a 2D harmonic  oscillator rapidly rotating at supercritical speed (i.e. above the natural normal mode frequency of oscillation). Its thermal noise is calculated in~\cite{PRLthermalnoise}, where the equations of motion  and their solution show that: $i)$ the signal determines the  equilibrium position and in  2D it is up-converted above the critical speed without attenuation; $ii)$ the offset errors by construction are reduced as the ratio spin-to-natural frequency squared (a well known phenomenon called self-centering); $iii)$ a weak instability (`whirl motion') occurs around the equilibrium position with time constant $\tau=\frac{Q}{\pi}P_{n}$, where $P_{n}$ is the natural period of oscillation and $Q$ the quality factor of the oscillator at the  (high) spin frequency not at the (low) natural one~\cite{Whirl1999}.

Measurements of the quality factor of CuBe suspensions by  Virgo scientists report  $Q=20000$ close to  $1\,\rm Hz$  (\cite{Qvirgo1999}, Fig.\,2). At a few Hz, for larger oscillation amplitudes at which higher losses are expected,  we have measured $Q=19000$~\cite{NobiliNASA1999}. With Q=20000,  for  a $1\,\rm Hz$ signal, thermal noise due to internal damping  in GG is only slightly larger than other sources of thermal noise, yielding altogether an integration time of $\simeq3.5\,\rm h$ for a signal-to-noise ratio SNR=2~\cite{IntegrationTimePRD2014}. Targeting a 100 times smaller signal requires an integration time $10^{4}$ times longer, hence it is quite remarkable  that, for the same SNR,  GG can reduce thermal noise below the signal  even quicker than Microscope (Table\,\ref{Tab:GGvsMicroscope}, entries IV, V and VI).  


Because of the limited duration of space missions --especially when they  rely on a finite amount of propellant to compensate solar radiation pressure and air drag-- a very short integration time is crucial  in order to ensure  a rigorous campaign of systematic  error checks. It needs a readout with low enough noise to match it.

\begin{figure}[htb]
\begin{center}
 \includegraphics[width=7.5cm,keepaspectratio]{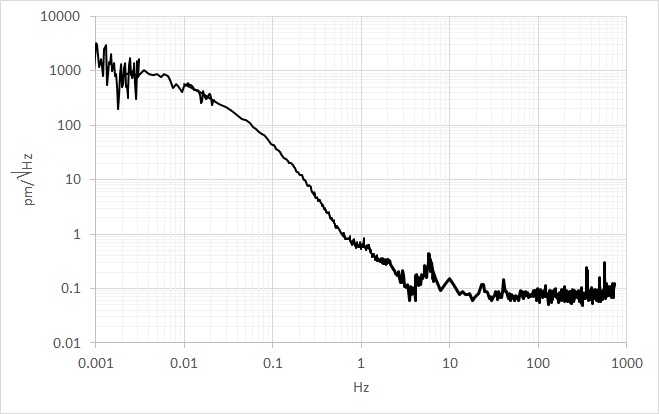}
     \caption{Displacement noise measured  at INRIM with a laser gauge suitable for GG~\cite{LIG2016}. At the GG signal frequency of $1\,\rm Hz$ the noise is  $0.6\,\rm\frac{pm}{\sqrt{Hz}}$. At high frequencies the noise measured is about   $0.1\frac{\rm pm}{\sqrt{\rm Hz}}$ (electronics noise, not interferometer noise). Noise at low frequencies is related to the optical fibers and can be reduced if needed. The measurement is not in vacuum and the frequency of the laser is not stabilized.}
\label{Fig:LIGdisplacementNoise}
\end{center}
 \end{figure}

In GG the differential displacements of the test cylinders are read by a heterodyne laser  interferometry gauge. It was originally proposed by Mike Shao, based on the laser gauge he developed at JPL for the SIM mission of NASA~\cite{Shao2010}. It is now under development at the Italian national metrology institute in Torino (INRIM)~\cite{LIG2015,LIG2016}, where it  has recently demonstrated  a displacement noise of $0.6\frac{\rm pm}{\sqrt{Hz}}$ at $1\,\rm Hz$ (Fig.\,\ref{Fig:LIGdisplacementNoise}), which means $1\,\rm s$ integration time for the target displacement of $0.6\, \rm pm$, up-converted to $1\,\rm Hz$. The high frequency of the signal makes the   GG interferometer much less complex than the one  which has successfully flown in LISA-PF (also a heterodyne laser  interferometer) because in this case low noise is required at frequencies well below $1\,\rm Hz$\,\cite{LISAPF2016}. For the GG laser gauge lab measurements have checked error sources such as laser frequency noise, cross talk and temperature sensitivity. A spurious effect which might arise because of rotation (the Sagnac effect) has been investigated, finding that it is much smaller than the signal~\cite{LIG2016}.

Comparison with  the capacitive readout of Microscope is shown in Table\,\ref{Tab:GGvsMicroscope}, entry VII (a cap readout is used also in the GG demonstrator on ground -- GGG).
By comparison with  capacitive sensors  the laser gauge is intrinsically differential, it does not require calibration, its sensitivity does not decrease with increasing gaps, it is less noisy. 

\clearpage
\begin{table}[h] \small \vspace{-5mm} 
\caption{Microscope and GG: a quantitative comparison}
\begin{tabular}{|>{\raggedright\arraybackslash}m{9cm}| >{\raggedright\arraybackslash}m{9cm}|  }

    \hline
    \hline
{\cellcolor[gray]{0.7}\hspace{3cm} \bf  Microscope} &
  { \cellcolor[gray]{0.7}\hspace{3.5cm} \bf GG} \\
\hline
\hline

    \hline
    \hline
    \multicolumn{2}{| c |}{\cellcolor[gray]{1} \bf I. Target of WEP violation $\eta=\Delta a/g(h)$, $g(h)\simeq8\,\rm ms^{-2}$} \\

$ \eta_{\mu scope}=10^{-15} $ &  
$ \eta_{_{GG}}=10^{-17} $\\
\hline
\hline

    \hline
    \hline
    \multicolumn{2}{| c |}{\cellcolor[gray]{1} \bf II. Signal: differential acceleration \& differential displacement} \\

\textit{Signal along symmetry axis of test cylinders (1D sensitivity)} & 
\textit{Signal in plane $\bot$ to symmetry/spin axis (2D sensitivity)} \\
 
$ \Delta a_{\mu scope}\simeq8\times10^{-15}\,\rm ms^{-2}$ & 
$ \Delta a_{_{GG}}\simeq8\times10^{-17}\,\rm ms^{-2}$  \\
 
$ \Delta x_{\mu scope}\simeq10^{-10}\,\rm m$ & 
$ \Delta r_{_{GG}}\simeq6\times10^{-13}\,\rm m$  \\

   \hline
    \hline


    \multicolumn{2}{| c |}{\cellcolor[gray]{1} \bf III. Test cylinders coupling, common mode rejection \& drag-free control} \\

 -- Each test cylinder individually suspended  by electrostatic forces along symmetry/sensitive axis, connected  to cage by Au wire& 
 -- Test cylinders coupled by CuBe flexures to form a beam balance with beam along symmetry/spin axis  sensitive in plane $\bot$ to it\\
 
 -- Common mode accelerations of individual cylinders to be reduced by matching their sensitivities with inflight calibrations:  ${\rm CMR}=18$ & 
 -- Beam balance to be balanced in flight (against common mode inertial acceleration due to drag) by adjusting balance arms using differential displacement as driver and piezo as actuators: ${\rm CMR}=10^{5}$  \\
 
  -- Drag-free control \& ${\rm CMR}=18$  will bring the residual differential acceleration due to drag (same frequency as signal) to $5.6\times10^{-15}\,{\rm ms^{2}}\simeq0.7\Delta a_{\mu scope}$ & 
  -- Drag-free control \& ${\rm CMR}=10^{5}$ will bring the residual differential acceleration due to drag (same frequency as signal) to $4\times10^{-17}\,{\rm ms^{2}}\simeq0.5 \Delta a_{_{GG}}$   \\

    \hline
    \hline
    
    \multicolumn{2}{| c |}{\cellcolor[gray]{1} \bf IV. Signal readout frequency} \\

 -- Inertial mode: displacement signal of  each test cylinder detected at  $\nu_{orb}\simeq1.7\times10^{-4}\,\rm Hz$  & 
 -- Differential displacement signal  detected at: $\nu_{spin}\simeq1\,\rm Hz$ \\
 
  -- Rotation mode ($\nu_{spin1}=3\nu_{orb}$): displacement signal of  each test cylinder detected at  $\nu_{1}=4\nu_{orb}\simeq5\times10^{-4}\,\rm Hz$  & 
  \\
  
    -- Rotation mode ($\nu_{spin2}=5\nu_{orb}$): displacement signal of  each test cylinder detected at  $\nu_{2}=6\nu_{orb}\simeq8.4\times10^{-4}\,\rm Hz$  & 
  \\

    \hline
    \hline
    \multicolumn{2}{| c |}{\cellcolor[gray]{1} \bf V. Internal damping at signal readout frequency} \\

$Q\simeq100$ & 
$Q\simeq20000$ \\

    \hline
    \hline
    \multicolumn{2}{| c |}{\cellcolor[gray]{1} \bf VI.  Integration time} \\

SNR=2 in $1.4\,\rm d$  integration time& 
SNR=2 in $3.5\,\rm h$ integration time \\

Typical  duration of science session: $8.3\,\rm d\ \ \rightarrow \ \   SNR=4.9$&
Typical duration of science session: $1\,\rm d\ \ \rightarrow \ \   SNR=5.2$\\

    \hline
    \hline
    \multicolumn{2}{| c |}{\cellcolor[gray]{1} \bf VII. Readout and readout noise} \\

 -- Capacitance readout for each test cylinder  ($600\,\mu\rm m$ gap )&
  -- Differential  heterodyne laser interferometer ($2\,\rm cm$ gap )  \\
 
 -- At $\nu\geqslant10^{-3}\,\rm Hz$~\cite{Touboul2009}:  & 
 -- At $\nu_{spin}\simeq1\,\rm Hz$: \\
 
 $38.5\,\rm\frac{pm}{\sqrt{Hz}}$ for outer cylinder displacement& 
 $0.6\,\rm\frac{pm}{\sqrt{Hz}}$  for relative displacements of test cylinders\\
 
  $40.4\,\rm\frac{pm}{\sqrt{Hz}}$ for inner cylinder displacement& 
\\

($\simeq200\,\rm\frac{pm}{\sqrt{Hz}}$ reported at  $10^{-3}\,\rm Hz$~\cite{Onera1999})&
\\

    \hline
    \hline
    \multicolumn{2}{| c |}{\cellcolor[gray]{1} \bf VIII. Radiometer non-gravitational effect} \\

-- Along sensitive axis  & 
-- Along axis $\bot$ to sensitive plane\\

-- Solved with: residual pressure of $10^{-5}\,\rm Pa$ \& passive temperature stabilization&
-- Causes indirect tidal effect in sensitive plane. Solved with: residual pressure of $10^{-5}\,\rm Pa$ \& passive  temperature stabilization\\

    \hline
    \hline
    \multicolumn{2}{| c |}{\cellcolor[gray]{1} \bf IX. Test cylinders centers of mass offsets} \\

-- $20\,\mu\rm m$ construction error for each test cylinder& 
-- $10\,\mu\rm m$ construction error for each test cylinder\\

-- $0.1\,\mu\rm m$ offset by inflight estimates \& \textit{a posteriori} data analysis& 
-- $0.6\,\rm nm$ self-centering of each test cylinder by physics  laws\\

  & 
 -- $1.7\,\rm nm$ test cylinders offset by capacitance control of whirl motion~\cite{Whirl1999}\\

    \hline
    \hline
    \multicolumn{2}{| c |}{\cellcolor[gray]{1} \bf X. Test cylinders instabilities \& control forces} \\

 %
   \begin{itemize}
   \setlength\itemsep{0.1em}
\item Each test cylinder (otherwise unstable)  controlled  along sensitive axis by capacitance sensors \& actuators; control always on~\cite{Rodrigues2009}
 \item  back action (DC)  for outer cylinder ($22.3\,\mu\rm m$ offset): $1.85\times10^{-9}\,\rm ms^{-2}$
 \item back action (DC)  for inner cylinder ($20.3\,\mu\rm m$ offset):  $7.64\times10^{-10}\,\rm ms^{-2}$
 \end{itemize} 
& 
 \begin{itemize}
   \setlength\itemsep{0.1em}
\item  Each test cylinder  weakly unstable by whirl motion with time constant $\tau=\frac{Q}{\pi}P_{n}\simeq9.6\,\rm d$  ($\rm Q\simeq20000$, $P_{n}=\frac{2\pi}{\omega_{n}}\simeq130\,\rm s$); controlled  by capacitance sensors \& actuators; control off during science data taking
 \item   control at whirl  frequency to keep offset $\leqslant1.7\,\rm nm$: $\geqslant4\times10^{-12}\,\rm ms^{-2}$
 \end{itemize} 
 \\
    \hline  
   \hline

    \end{tabular}
\label{Tab:GGvsMicroscope}
    \end{table}
\clearpage


Bringing the radiometer effect 100 times below the Microscope target is impossible at room temperature (in the Stanford proposal cryogenics  would ensure  an extremely low pressure which makes radiometer negligible). Radiometer acts along the symmetry axis.  GG can deal with it at room temperature because it is  sensitive in the plane perpendicular to it. 
Nonetheless,  the largest tidal effect between the test cylinders of GG in the sensitive plane is indirectly due to radiometer along the much stiffer symmetry axis (Table\,\ref{Tab:GGvsMicroscope}, entry VIII).

Direct tidal effects come from offsets between the centers of mass of the test cylinders in the sensitive plane. Self-centering as predicted theoretically ensures, for the test cylinders of GG, a reduction of the offsets achieved by fabrication by more than 4 orders of magnitude (Table\,\ref{Tab:GGvsMicroscope}, entry IX; see~\cite{IJMPD}, Fig. 5 for experimental evidence of self-centering with the  GGG demonstrator). 

As shown in~\cite{PRLthermalnoise},  losses generate a  whirl motion around the equilibrium position at the natural frequency.  
 It is known that, except in the presence of large dissipation,  the frequency of whirl is the same as the natural frequency at zero spin. However, there was no experimental demonstration so far  that the relevant quality factor is that at the spin frequency, not  at the natural one. Since losses are lower at higher frequencies, the issue is very important. 

\begin{figure}[htb]
\begin{center}
 \includegraphics[width=7.5cm,keepaspectratio]{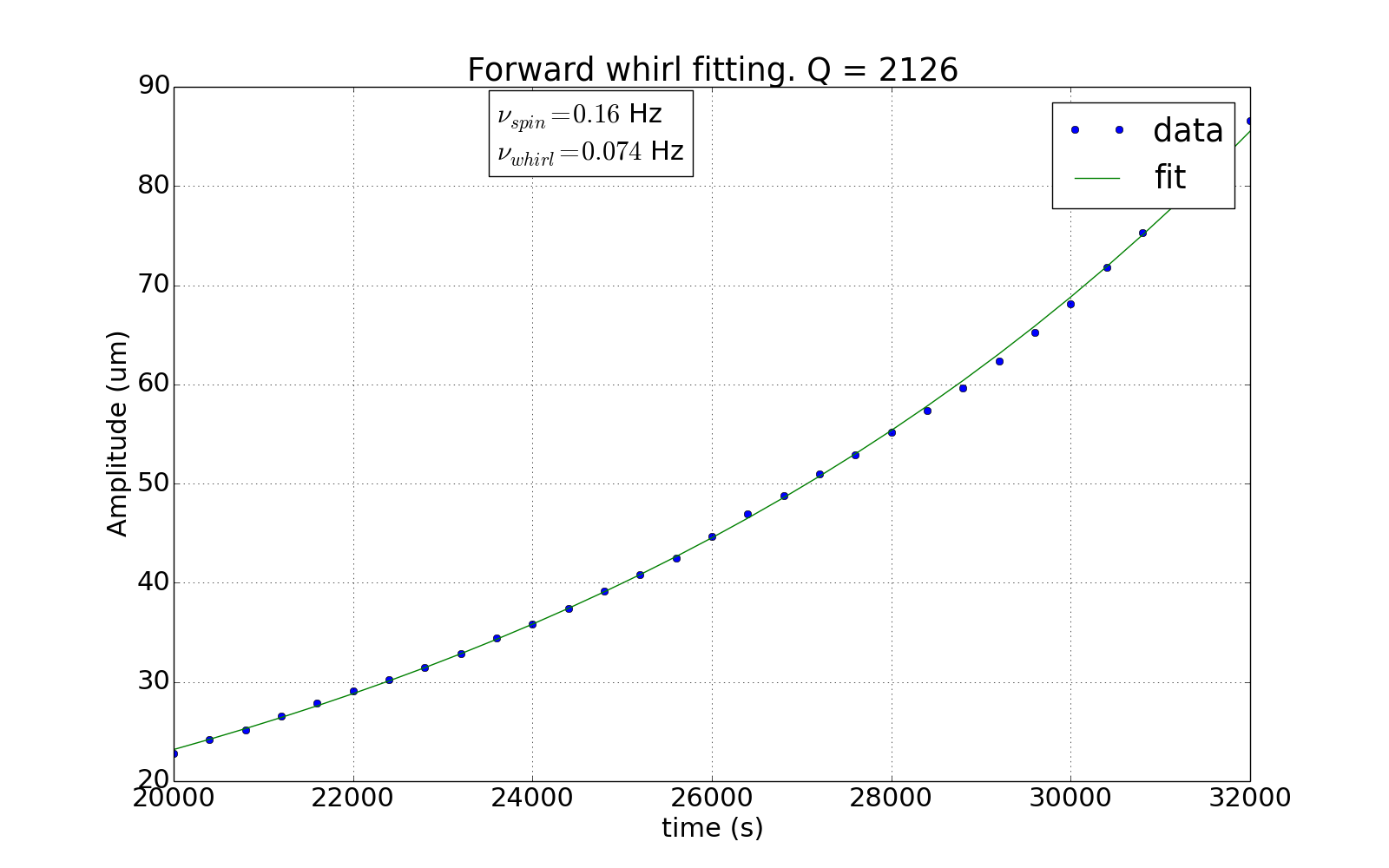}
     \caption{Exponential growth of  whirl motion of the test cylinders relative to each other  in the GGG laboratory demonstrator with time constant $\tau=\frac{Q}{\pi}P_{whirl}$. The system spins at $\nu_{spin}=0.16\,\rm Hz$ and the whirl period is  $P_{whirl}=\frac{2\pi}{\nu_{whirl}}=13.5\,\rm s$. We measure:  $Q_{whirl}=2126$, hence $\tau\simeq2.5\,\rm h$.}
\label{Fig:QwhirlRotante}
\end{center}
 \end{figure}

\begin{figure}[htb]
\begin{center}
 \includegraphics[width=7.5cm,keepaspectratio]{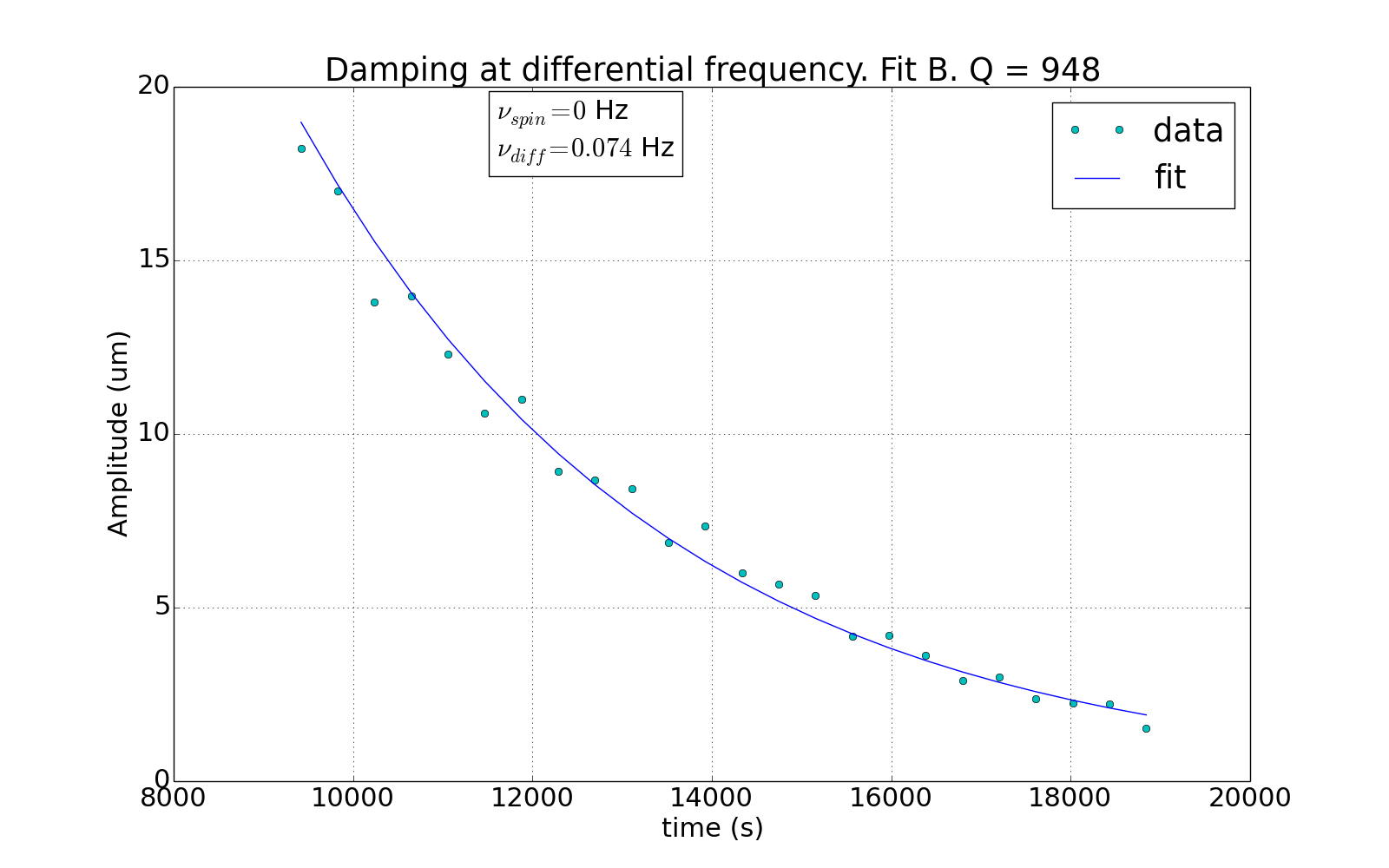}
     \caption{
     Exponential decay of the relative  oscillation amplitude of  the GGG test cylinders  not rotating and oscillating  at the natural differential mode frequency $\nu_{diff}=0.074\,\rm Hz$.
     The system is the same as in Fig.\,\ref{Fig:QwhirlRotante} after stopping rotation and before breaking  vacuum.
    We measure  $ Q_{\nu_{diff}}=948$ (with a decay time constant of $1.1\,\rm h$). 
     }  
\label{Fig:QwhirlNonRotante}
\end{center}
 \end{figure}

  The GGG demonstrator, with  complex CuBe cardanic joints  to sustain weight, a spin frequency  $\nu_{spin}=0.16\,\rm Hz$ and a differential mode frequency of $\nu_{diff}=0.074\,\rm Hz$ ($P_{diff}=13.5\,\rm s$), shows that  whirl grows with $Q_{\nu_{spin}}=2126$ and  time constant $\tau=\frac{P_{diff}}{\pi}\,Q_{\nu_{spin}}\simeq2.5\,\rm h$ (Fig.\,\ref{Fig:QwhirlRotante}), while at zero spin, the amplitude of oscillation at the differential mode frequency decays faster, with $Q_{\nu_{diff}}=948$ (Fig.\,\ref{Fig:QwhirlNonRotante}).

 The theoretical predictions are  demonstrated: the frequency of whirl is the same as the natural differential frequency,  whirl growth occurs as in the analytical solution and the quality factor  improves when the system spins faster than this frequency. Nonetheless, at $1$-$g$ it would be hard to have a time constant of whirl growth longer than $2.5\,\rm h$ (a long differential period requires weak suspensions, which is impossible with $10\,\rm kg$ test cylinders at $1$-$g$), and this means that in GGG whirl control, performed by means  of capacitance sensors/actuators,  must be on all the time (see~\cite{GGfocusIssue2012} for the results of  one month measurements).

In space with lower stiffness, hence lower natural frequencies, higher spin rate hence higher quality factor, whirl growth is much slower, with a much longer  time constant. In GG the time constant for the whirl growth of each test mass is almost $10$ days, whirl damping (performed with capacitors as on ground) requires very weak forces and it is  off during science data taking, thus leaving the test masses totally  passive, as  advocated by Braginsky (Table\,\ref{Tab:GGvsMicroscope}, entry X).  On ground, only rotors with high rotor/stator  and bearings noise, or very high dissipation have reported chaotic behaviour~\cite{Chaos1,Chaos2,Chaos3}. 

\clearpage


 In such rotors  the frequency of whirl when the system rotates is no longer equal to the corresponding  natural normal mode frequency at zero spin, which is `the smoking gun'  for the onset  of whirl chaos. 
GG has no motor, no bearings,  no stator and high $Q$, hence whirl chaos is ruled out. Experimental evidence is provided by the GGG  demonstrator on ground, which has motor, stator and bearings noise and  higher dissipation than in space (due to more complex suspensions and a lower spin rate), and yet  the whirl and the normal mode frequencies have always been found to be the same.


%
 %

 
The spin axis of GG (perpendicular to the sensitive plane) is essentially undisturbed and remains  fixed in space due to its high rotation energy. Instead, being in a sun-synchronous orbit, the  normal to the orbit plane precesses around the north pole by $\simeq1^{\circ}/\rm d$. Over an orbital arc of $90\,\rm d$, in a totally passive and deterministic manner, the angle between the two varies from $-45^\circ$ to $+45^\circ$, thus making the  violation signal vary in a perfectly known way, different from all systematic effects --even at the same frequency-- that we are aware of.  With  $90$ measurements during the cycle (Table\,\ref{Tab:GGvsMicroscope}, entry VI),  all to the target precision,  the variation of the measured effect with the changing geometry is mapped and the signal, if present, is separated from errors with certainty (there will be at least three  $90$-d cycles during the mission). A second balance with equal composition test cylinders, co-axial and concentric with the one with different composition cylinders, can also be accommodated for further checking~\cite{GG4TestMasses2003}.


 The test cylinders of GG spin around their symmetry axis. Manufacturing imperfections, being fixed with the cylinders, produce DC effects on their relative displacements which do not compete with the much higher frequency of the signal.  This unique feature, together with large test masses ($10\,\rm kg$ each), self-centering and balancing,  allows  requirements on  their fabrication to be relaxed. It is therefore possible to test `unusual' materials which are known to probe WEP violation  more deeply but have never been implemented because of fabrication issues.  The choice of $\rm Pb$ and $\rm C_{2}H_{4}$ proposed for GG in 2009 in~\cite{GGphaseA22009} has been found to yield --with the same driving signal and the same sensitivity to differential accelerations-- a better test of WEP by a factor of 12~\cite{MikePRL2013},  which would mean the possibility of reaching $10^{-18}$. 

We have discussed the key facts whereby  GG can aim at a WEP test to $10^{-17}$. Although a full report on  the error budget (and consequent  experiment and mission requirements) is beyond the scope of this paper, it is worth listing the major systematic errors  which compete with a WEP violation signal at this level. As shown in Table\,\ref{Tab:ErrorBudget}, these are, in the first place, the effects that have the same frequency as the signal (in the non rotating frame of the spacecraft this is the orbital frequency $\nu_{orb}\simeq1.7\times10^{-4}\,\rm Hz$, which  the rotation of the s/c   up-converts to $1\,\rm Hz$ as shown in Fig.\,\ref{Fig:GGsignalAndUpconversion}) and  secondly, those not too far from the signal frequency, at twice it. As expected, the most relevant error is due to non-gravitational forces on the outer surface of the spacecraft (although at a large phase difference from the signal)  which GG  can reduce below the target  only thanks to the unique feature of rejecting common modes in flight. Frequency separation allows effects at $2\nu_{orb}$ which  are larger than the signal   to be clearly distinguished  from it. 
Effects at frequencies much farther away from $\nu_{orb}$, such as the natural/whirl frequencies, can be  much bigger than the signal and still be clearly distinguished from it.

GG has been under investigation for over $20$ years (including the construction and testing of a  full scale ground demonstrator). The most comprehensive industrial study funded by ASI (Agenzia Spaziale Italiana), reported  in about 30 documents~\cite{GGPRRASI2009},   features a full 3D simulation of the space experiment based on the simulator developed at Thales Alenia Space in Torino, Italy for the successful GOCE mission of ESA~\cite{GOCEgiuseppe}.

\begin{table}[h!] \small \vspace{-6mm} 
\caption{Budget of the major systematic errors for GG \\ targeting a WEP test to  $ \eta=10^{-17}$}
\begin{tabular}{|>{\raggedright\arraybackslash}m{4cm}| >{\raggedright\arraybackslash}m{1.8cm}| >{\raggedright\arraybackslash}m{2cm}|   }
    
    \hline
    \hline
\cellcolor[gray]{0.9}\textbf{Acceleration ($x,y$ sensitive plane) due to:}&
\cellcolor[gray]{0.9}\textbf{Frequency}&
   \cellcolor[gray]{0.9}   Differential acceleration $\rm \bf{(m/s^2)}$ 
\\
\hline
\hline

\textbf{WEP violation signal }&
$ \nu_{orb} $ & 
$ 8.1\cdot10^{-17} $
\\ 
\hline
\hline 
\hline
\hline

External non-gravitational forces after drag compensation {\underline{and}} common mode rejection&
$ \nu_{orb} $   & 
$ 4\cdot10^{-17} $
\\ 
\hline
\hline 
 
 Earth's  monopole coupling with test masses quadrupole moments&
$ \nu_{orb} $  & 
$ 6\cdot10^{-18} $
\\ 
\hline
  
Earth's tide coupled to radiometer effect along spin axis  $ z $ &
$ \nu_{orb} $   & 
$ 4\cdot10^{-18} $
\\ 
\hline
 
 Earth's  tide coupled to thermal emission along spin axis  $ z $ &
$ \nu_{orb} $   & 
$ 8.1\cdot10^{-19} $
\\ 
\hline

Earth's tide coupled to non-gravitational accelerations along spin axis  $ z $ &
$ \nu_{orb} $   & 
$ 1.7\cdot10^{-19} $
\\ 
\hline

Earth's tide coupled to radiometer effect along spin axis  $ z $ &
$ 2\nu_{orb} $   & 
$ 8.5\cdot10^{-16} $
\\ 
\hline

Earth's tide coupled to non-gravitational accelerations along spin axis  $ z $ &
$2\nu_{orb} $   & 
$ 3.4\cdot10^{-17} $
\\ 
\hline

Magnetic dipole moment of one test mass coupled to magnetization induced on the other  by Earth's magnetic field $B_{\oplus}$&
$ 2\nu_{orb} $   & 
$ 1.8\cdot10^{-16} $
\\ 
\hline

   \end{tabular}
\label{Tab:ErrorBudget}
    \end{table}


\section{Conclusions}
\label{Sec:Conclusions}

In this paper we have shown how the progress in WEP tests  has depended on innovations in experimental techniques, driven by theoretical insight. 

The torsion balance, first used to test the equivalence of inertial and gravitational mass at the end of the 19th century,  was the first high accuracy instrument: intrinsically differential to reject common mode forces,  coupled to the lab by a thin fiber with very low torsional stiffness   to enhance sensitivity and high mechanical quality for low thermal noise, it was limited by  the lack of modulation of the signal in the  field of Earth. Later on it was realized that, by  taking the Sun as source, the  diurnal rotation of  Earth would provide  the desired modulation, with no need to flip or rotate the balance. Next,  the ability to smoothly rotate the torsion balance itself,  and to reduce and/or compensate gravity gradients, along with  the   understanding that losses due to internal damping decrease at higher frequency, whereby  a rotating  balance moves the signal to a region where thermal noise is lower, have led to about five orders of magnitude improvement in the field of the Earth and almost one in the field of the Sun. 

Eventually, rotating torsion balance experiments are limited by gravity gradients, particularly those changing with time, and by thermal noise driven by losses in the suspension fiber at the rotation frequency of the balance. Current laboratory experiments have reached a point where significant improvement is hard to achieve, and the same applies to LLR, for the reasons discussed in Sec.\,\ref{Sec:StateOfArt}. 

Advancing by orders of magnitude requires moving the experiment to space.

Microscope is the first WEP experiment to have been implemented in space, realizing a concept originally  proposed (in a cryogenic version) at Stanford University in the 1970s. The Microscope sensor is made of two concentric cylinders --their common symmetry/sensitive  axis lying in the plane of the orbit-- whose  displacements along the axis  are individually read and controlled in order to minimize the  separation between their centers of mass. In low Earth orbit the driving acceleration is almost 500 times larger than for a torsion balance in the lab, and by this fact alone a corresponding increase in sensitivity can be targeted. The main design parameters of Microscope are discussed in Sec.\,\ref{Sec:Microscope}. It appears that the $10^{-15}$ target  would be very hard to improve with this design because of such features as  tidal  accelerations resulting from relative position errors by construction and mounting, residual non-gravitational  accelerations due to poor common mode rejection, large control forces to be applied in the direction of the signal,  thermal noise due to large losses at low frequency in a loose gold wire. Even cryogenics, with all the attendant practical limitations and problems, would not solve all the open issues.

The GG sensor, too, is a pair of concentric cylinders but their symmetry axis is placed orthogonal to, rather than in, the orbit plane, and is sensitive in 2D rather than along the axis. By this simple choice, the rotation rate is no longer limited by stability issues and can be chosen sufficiently high where  losses due to internal damping are low and   thermal noise is small. With carefully designed U-flexures and some ingenuity, the two  concentric cylinders are coupled as in a beam balance with the beam along the symmetry/rotation axis,  sensitive in the plane perpendicular to it.  
In flight, against the common mode effect of drag which is many orders of magnitude weaker than $1$-$g$, the beam balance  can be precisely balanced by adjusting its arms,  more easily  than a  balance on ground against the very large effect of local gravity, thus  achieving good common mode rejection.
 As a result, air drag is  partially  compensated by drag-free control and partially rejected by the balance, making its residual differential effect much smaller than it could possibly be  by active compensation only, with no need to push the requirement for low noise  trusters  to  unrealistic limits.
When rotating above a critical rate, physics itself takes care of aligning the rotation axis to the symmetry axis much better than it can possibly be achieved by construction (self-centering), the remaining offset being very small, calculable and measurable with the ground demonstrator.  Because of losses  in the flexures, motion around the equilibrium position  is weakly unstable (whirl motion). The relevant losses which determine the growth rate are predicted to occur at the (high) spin frequency (at which losses are known to be small). This fact, predicted  theoretically and demonstrated experimentally  (see Sec.\,\ref{Sec:GG})  ensures a very slow growth, such that in GG whirl control can be switched off during science data taking, leaving the test cylinders totally passive.

GG was conceived to address and solve the shortcomings of alternative experiment designs, as they have emerged in an almost 45-year long history of proposals for testing the WEP in space. A comparison of the key issues, as  carried out in Secs.\,\ref{Sec:Microscope} and \ref{Sec:GG} and summarized in Table\,\ref{Tab:GGvsMicroscope}, shows how the GG approach makes it possible to target a WEP test 100 times more precise than Microscope, to $10^{-17}$. Should Microscope detect a sign of new physics, another mission capable of greater precision would be called for, and urgently. As we show in Sec.\,\ref{Sec:GG}, a design based on different principles is needed. 


In recent years, three space missions have shown that very high precision physics experiments are feasible in space, and that the relevant technological challenges can be met. Most technological ingredients of GG have received confirmation as part of successful satellite missions. GOCE (2009-2013) demonstrated drag-free control,  high passive thermal stability and high precision capacitive measurement and control technology. LISA-PF (2015-2017) confirmed these findings and demonstrated very low noise heterodyne laser metrology on cm-length scales. Microscope (2016-ongoing) showed that precise tests of  WEP in space are feasible and  can take advantage of rotation because --contrary to the widespread concern, originating from lab experience, whereby rotation impairs precision--  the space environment is favorable to rotation for the simple reason that in space, unlike on ground, the whole `lab' (the spacecraft) rotates relative to inertial space.  As to the GG instrument, its high quality suspensions have heritage from gravitational wave detectors and the  \eotwash\  balance, and it has been  tested as far as feasible for an instrument designed to fly  with a dedicated demonstrator at 1-$g$.

Currently, GG is a candidate in the ESA M5 competition, the final phase of which will take place in the same time frame as the disclosure of the early results of Microscope. Should the need arise, peer review of all proposed experiments  would be a prerequisite before a confirmation mission is launched.  The arguments reported here indicate that GG has good chances to emerge as the  most viable and cost-effective option.

\vspace{0.5cm}

 \textbf{Acknowledgements.}  Thanks are due to all colleagues of the GG collaboration, especially for contributing to the proposal submitted in response to the ESA Call for the next medium size mission M5.  The GG project has received support from ASI for space mission studies and the GGG laboratory prototype, also funded by INFN. The support of ESA for the development of a low noise laser gauge is gratefully acknowledged.  


\end{document}